\documentclass[11pt]{article}
\usepackage{geometry}               
\usepackage{slashed}
\usepackage{graphicx}
\usepackage{amsmath}
\usepackage{amssymb}
\usepackage{epstopdf}
\usepackage{verbatim}
\usepackage{url}
\usepackage{layout}
\usepackage{setspace}
\def \tk{\tilde k}

\newcommand{\reef}[1]{(\ref{#1})}

\newcommand{\Z}{\mathbb{Z}}

\newcommand{\sqNMHV}{{\rm \sqrt{N}MHV}}
\newcommand{\dash}{\text{-}}

\newcommand{\cn}{{\cal N}}

\newcommand{\be}{\begin{equation}}
\newcommand{\ee}{\end{equation}}

\def\be{\begin{equation}}
\def\ee{\end{equation}}
\def\bea{\begin{eqnarray}}
\def\eea{\end{eqnarray}}
\def\ba{\begin{array}}
\def\ea{\end{array}}
\def\bd{\begin{displaymath}}
\def\ed{\end{displaymath}}

\def\ie{{\it i.e.~}}

\def\a{\alpha}

\def\l{\lambda}
\def\m{\mu}

\def\>{\rangle} %right angle
\def\<{\langle} %left angle
\def\Dsl{D \hskip-.6em \raise1pt\hbox{$ / $ } }
\def\to{\rightarrow}

\def\lab{\label}

\newcommand{\eps}{\epsilon}
\newcommand{\lra}{\leftrightarrow}

\begin{document}

\begin{flushright}
MCTP-10-25 \\
PUPT-2345 \\
\end{flushright}
\vspace{3.0cm}

\begin{center}
{\Large \bf Stringy KLT relations, global symmetries,
and $E_{7(7)}$-violation}

\vspace{0.2in}
{\bf Henriette Elvang$^{a,b}$ and Michael Kiermaier$^c$}\\
\vspace{0.2cm}
{\it $^a$Michigan Center for Theoretical Physics (MCTP)\\Department of Physics, University of Michigan, \\ Ann Arbor, MI 48109}\\
\vspace{0.2cm}
{\it $^b$Institute for Advanced Study\\ Princeton, NJ 08540}\\
\vspace{0.2cm}
{\it $^c$Department of Physics, Princeton University \\ Princeton, NJ 08544}\\
\end{center}

\vspace{0.4cm}
\centerline{\small Email: {\tt elvang@umich.edu, mkiermai@princeton.edu}}
\vspace{0.4cm}

\begin{abstract}
We study consequences of the Kawai-Lewellen-Tye (KLT) relations applied to tree amplitudes in toroidal compactifications of string theory to four dimensions. The closed string tree amplitudes with massless external states respect a global $SU(4) \!\times\! SU(4)$ symmetry, which is enhanced to the $SU(8)$ $R$-symmetry of  $\cn=8$ supergravity in the field theory limit. Our analysis focuses on two aspects: (i) We provide a detailed account of the simplest $SU(8)$-violating amplitudes. We classify these processes and derive explicit superamplitudes for all  local $5$- and $6$-point operators with $SU(4) \!\times\! SU(4)$ symmetry at order $\alpha'^3$.
Their origin is the dilatonic operator $e^{-6 \phi} R^4$ in the closed-string effective action.
(ii) We expand the 6-point closed string tree amplitudes to order $\alpha'^3$ and use two different methods to isolate the $SU(8)$-singlet contribution from $e^{-6 \phi} R^4$. This allows us to extract the matrix elements of the unique $SU(8)$-invariant supersymmetrization of $R^4$. Their single-soft scalar limits are non-vanishing. This demonstrates that the $\cn=8$ supergravity candidate counterterm $R^4$ is incompatible with continuous $E_{7(7)}$ symmetry.
From the soft scalar limits, we reconstruct to quadratic order the $SU(8)$-invariant function of scalars that multiplies $R^4$, and show that it satisfies the Laplace eigenvalue equation derived recently from supersymmetry and duality constraints.

\end{abstract}

\newpage

\setstretch{0.3}
\tableofcontents
\setstretch{1.05}
\newpage

\section{Introduction}
In this paper we present consequences of the  Kawai-Lewellen-Tye (KLT) relations~\cite{KLT} in string theory toroidally compactified to four dimensions.
The KLT relations encode a  remarkable relationship between $n$-point closed- and open string tree amplitudes. In the field-theory limit, $\alpha'\to0$, the relations are often expressed as ``\,gravity = (gauge theory)$^2$\,''. 
Schematically, we write KLT as
\bea\label{KLTschematic}
  M_n = \sum_\mathcal{P} f_n(s_{ij}) \, A_n \times  \tilde{A}_n \, ,
\eea
where $M_n$ is a closed string tree amplitude, $f_n$ is a function of Mandelstam variables, and $A_n$ and $\tilde{A}_n$ are ordered tree amplitudes in the ``left'' and ``right'' open string theories. The sum is over certain permutations $\mathcal{P}$ of external states.
Our study is restricted to string tree amplitudes with external states exclusively from the massless sectors, i.e.~the states of $\cn=4$ SYM and $\cn=8$ supergravity.

Closed string theory does not have any continuous global symmetries; however, the tree level amplitudes with massless external states vanish unless the states form an $SU(4) \times SU(4)$ invariant. The origin of this symmetry is the $SO(6,6)$ duality group of the toroidal compactification. The $SU(8)$ $R$-symmetry group of $\cn=8$ supergravity is broken explicitly to  $SU(4)\times SU(4)$ by the $\alpha'^3$-corrections to the closed string 
 tree 
amplitude.
In the  effective action this can be understood through the appearance of the operator $e^{-6\phi}R^4$, where $\phi$ is the dilaton. We construct $SU(8)$-violating amplitudes explicitly via KLT and study their supersymmetrizations. This is naturally connected to the promotion of KLT to \emph{super-KLT relations} for superamplitudes.

In $\cn=8$ supergravity, any possible logarithmic divergences 
 in 
loop computations~\cite{BernFiniteness} based on generalized unitarity~\cite{Unitarity}  correspond to supersymmetric and $SU(8)$-invariant counterterms.
Operators such as $D^{2k}R^4$ are studied as candidate counterterms~\cite{Counterterms,Harmonic,StringyCounterterms,Kallosh,Eisenstein,efk5} (see~\cite{LanceReview} for a recent review). The  global $E_{7(7)}$ symmetry~\cite{N8} of $\cn=8$ supergravity is spontaneously broken to $SU(8)$, and  the 70 scalars $\varphi$ of $\mathcal{N}=8$ supergravity are the Goldstone bosons. $E_{7(7)}$ requires that the matrix elements of a candidate counterterm vanish in the single-soft scalar  limit (for recent discussions, see~\cite{BEF,ArkaniHamed:2008gz,footprint}). In this paper, we
show that the 6-point matrix elements of the 3-loop counterterm $R^4$ have non-vanishing single-soft scalar limits; therefore $R^4$ violates $E_{7(7)}$. We use the string tree amplitudes as a technical tool for computing the matrix elements of $R^4$. This requires that we remove the $SU(8)$-violating contributions from string moduli in order to isolate the $SU(8)$-invariant part, and we develop the tools to do so.

%he edits paragraph below
As noted above, the closed string effective interaction $e^{-6\phi}R^4$ violates $SU(8)$. Let us present a simple direct illustration of this. The massless $\cn=8$ closed string spectrum has 35 pairs of complex scalars, which decompose in three different ways into products of open string states: 1 pair $\varphi$, $\bar{\varphi}$ from opposite helicity gluons, 16 pairs $\varphi_f$, $\bar{\varphi}_f$ from opposite helicity gluinos, and 18 pairs $\varphi_s$, $\bar{\varphi}_s$ from scalars. In the field theory limit $\alpha'\to 0$, the 35 scalar pairs are $SU(8)$-equivalent. However, at order $\alpha'^3$, the 3 different 6-point matrix elements
with two conjugate scalars and four gravitons are distinct; in particular, they have inequivalent single-soft scalar limits. An explicit KLT computation, using the open string tree amplitudes of~\cite{STMHV,ST6pt} as input, gives
\begin{eqnarray}
  \nonumber
    \!\!\!\!&&\lim_{p_1\to0} \bigl\<\,\varphi\,\,  \bar{\varphi}\,\,++--\bigr\>_{e^{\dash6\phi}\!R^4} ~=~6\times[34]^4\<56\>^4\,,\quad
    \lim_{p_1\to0} \bigl\<\varphi_f \bar{\varphi}_f++--\bigr\>_{e^{\dash6\phi}\!R^4} ~=~3\times[34]^4\<56\>^4\,,\\[1mm]
    \!\!\!\!&&\lim_{p_1\to0} \bigl\<\varphi_s  \bar{\varphi}_s++--\bigr\>_{e^{\dash6\phi}\!R^4} ~=~0\,.\label{softe6fR4}
\end{eqnarray}
If $SU(8)$ were unbroken, these matrix elements (and their soft scalar limits) would have been identical. The fact that they are different is a clear signature of $SU(8)$-violation.
Let us note that the result (\ref{softe6fR4}) shows that some matrix elements of $e^{-6\phi}R^4$ exhibit non-vanishing single-soft scalar limits, consistent with the results of a previous analogous computation of Br\"odel and Dixon~\cite{BD}. Since $e^{-6\phi}R^4$ violates $SU(8)$, the maximal compact subgroup of $E_{7(7)}$, there is no reason for the scalar soft-limits to vanish. Note, however, that the third soft limit in~(\ref{softe6fR4}) vanishes. This is no coincidence: when $\a'>0$, $E_{7(7)}$ is explicitly broken to $SO(6,6)$, and the scalars $\varphi_s\,  \bar{\varphi}_s$ are the 36 Goldstone bosons of the spontaneous breaking of $SO(6,6)$  to $SU(4) \times SU(4)$.\footnote{We have verified $\lim_{p_1\to0} \bigl\<\varphi_s\,  \bar{\varphi}_s++--\bigr\> ~=~0$ for the closed string amplitude up to and including order $\a'^7$.} 
%he footnote added

We use two different methods to extract the $SU(8)$-invariant part of the supersymmetrization of $e^{-6\phi}R^4$. The resulting matrix elements are the matrix elements of the unique
$SU(8)$-invariant 
non-linear completion 
of $R^4$. In the first method, we
obtain the matrix elements of $R^4$ by ``averaging'' the matrix elements of $e^{-6\phi}R^4$ over $SU(8)$.
This approach is presented in section \ref{s:SSLgrp} and requires no knowledge of superamplitudes.
The second method (section \ref{s:SSL}) applies the superamplitude technology developed in sections~\ref{s:superA} and~\ref{s:local} (extending earlier work in~\cite{efk4,efk5}) to systematically isolate and remove the $SU(8)$-violating contribution. In both approaches we find that the $6$-point NMHV matrix elements of $R^4$ have non-vanishing single-soft scalar limits, specifically
%he added phi_f's also
\begin{equation}
  \label{softR4}
    \lim_{p_1\to0} \bigl\<\varphi\,  \bar{\varphi}++--\bigr\>_{R^4} ~=~
    \lim_{p_1\to0} \bigl\<\varphi_f\,  \bar{\varphi}_f++--\bigr\>_{R^4}
     ~=~
    \lim_{p_1\to0} \bigl\<\varphi_s\,  \bar{\varphi}_s++--\bigr\>_{R^4} ~=~-\frac{6}{5}\,[34]^4\<56\>^4\,.
\end{equation}
The difference between~(\ref{softe6fR4}) and~(\ref{softR4}) can be attributed precisely to the $SU(8)$-violating pollution from 
$\phi^2R^4$. 

We emphasize that the matrix elements of the $SU(8)$-invariant supersymmetrization of $R^4$ are unique. If there were a second non-linear completion of $R^4$ respecting $SU(8)$, the difference between both completions would be a local supersymmetric dimension-8 operator.\footnote{
In this work, we define the dimension of an operator as the mass dimension of its matrix elements with the couplings ($\kappa$ and $\a'$) stripped off. In this counting, all terms in the non-linear completion of an operator have the same dimension, and all $L$-loop candidate counterterms of $\cn=8$ supergravity have dimension $2L+2$.} Its leading matrix elements are $5$-point or higher, because both completions of $R^4$ must coincide at $4$-point. However, no such higher-point dimension-8 operator exists~\cite{Harmonic,efk5}. We conclude that the $E_{7(7)}$-incompatibility of the soft limits~(\ref{softR4}) cannot be cured by any non-linear modification of $R^4$.

Our results are consistent with the findings of Green, Miller, Russo and Vanhove~\cite{Eisenstein},  
whose analysis of moduli-dependence showed that $R^4$ must necessarily violate $E_{7(7)}$-symmetry. Specifically, these authors derive constraints from duality and supersymmetry on the moduli-dependence of operators $f(\varphi)R^4$. In four dimensions,  the constraint takes the form of an eigenvalue equation  $\Delta f(\varphi)=\lambda f(\varphi)$, where $\Delta$ is the Laplacian of the coset-space $E_{7(7)}/SU(8)$ parameterized by the 70 $\varphi$'s. We show that the coefficient $-6/5$ in~(\ref{softR4}) is precisely the one needed to satisfy the Laplace equation at quadratic order in the scalars.
The analysis of~\cite{Eisenstein} suggests that the operators $D^4R^4$ and $D^6R^4$ are also incompatible with $E_{7(7)}$.
It would be interesting to extend our analysis to study these $5$- and $6$-loop candidate counterterms.

Motivated by the $SU(4)\times SU(4)$ symmetry of closed string tree amplitudes, we present a detailed treatment of superamplitudes and local operators with this symmetry. The closed string tree amplitudes do not obey the simple N$^k$MHV amplitude classification of supergravity, but are instead labeled by  two integers $k$ and $\tilde{k}$: in KLT, an N$^{(k,\tilde{k})}$MHV closed string tree amplitude stems from the product of an N$^k$MHV open string amplitude $A_n$ and an N$^{\tilde{k}}$MHV open string amplitude $\tilde{A}_n$.
We find that without $SU(8)$, $R^4$ is no longer the unique dimension-8 operator, but instead there are additional, independent local supersymmetric operators with $n>4$ fields. These can be understood as dilatonic operators of the form $\phi^m R^4$. We present explicit expressions for the superamplitudes associated with $\phi R^4$ and $\phi^2 R^4$.

Recently, BCFW recursion relations~\cite{bcfw} have been used to derive quadratic relations between gauge-theory tree amplitudes~\cite{BDFS}. These identities can be understood from KLT as the statement that $SU(8)$-violating amplitudes must vanish in the $\alpha'\to0$ limit. Aspects of this were also discussed recently in \cite{Feng,Tye}. We present generalizations of these identities, in particular new {\em linear} relations between the matrix elements of the supersymmetrization of $F^4$. These relations immediately follow from the vanishing of the closed string tree amplitude at order $\alpha'^2$, which was also noted by~\cite{BD}.

\smallskip
{\em The paper is structured as follows.}
In section \ref{2} we introduce the spectrum of massless states, the KLT relations and the relevant symmetries. We then present the simplest example of an $SU(8)$-violating amplitude. Section \ref{s:SSLgrp} contains our first approach to isolating the $SU(8)$-invariant matrix elements of $R^4$ from the $O(\alpha'^3)$ string tree amplitude, and we compare their non-vanishing single-soft scalar  limits to the constraints derived in~\cite{Eisenstein}.  We introduce the N$^{(k,\tilde{k})}$MHV classification of the closed string tree amplitudes in section \ref{s:superA} and illustrate how super-KLT relations work. We present $SU(4)\times SU(4)$ preserving superamplitudes and solve their SUSY Ward identities in the simplest sectors. In section~\ref{s:local}, we construct $5$- and $6$-point superamplitudes for $SU(8)$-violating local operators of dimension $8$.
The superamplitude technology is then used in section \ref{s:SSL} as our second method to show that $R^4$ is not $E_{7(7)}$ compatible, and we determine the explicit superamplitude that relates the $6$-point matrix elements of $e^{-6\phi}R^4$ and $R^4$. In section \ref{s:vanishing}, we present new linear identities for the matrix elements of $F^4$.

\smallskip
The closed string tree amplitudes in this work are all obtained from open string tree amplitudes via KLT. We are grateful to Stieberger and Taylor for having provided us with Mathematica codes containing the $\alpha'$-expansions of the NMHV open string tree amplitudes of their paper \cite{ST6pt}.

%%%%%%%%%%%%%
\setcounter{equation}{0}
\section{KLT relations and $SU(8)$-violating processes}\label{2}

\subsection{Spectrum, KLT and symmetries}
We study string amplitudes with massless external states of  string theory, toroidally compactified to four dimensions. The external states in the open string amplitudes will be the 16 massless states of $\mathcal{N}=4$ SYM. The gluons, four gluinos and six scalars transform in $r$-index antisymmetric representations of $SU(4)$ for $r=0,\dots, 4$, i.e.~the gluons are singlets, the gluinos transform in the {\bf 4} and $\overline{\bf 4}$, and the scalars  in the {\bf 6}. We label them as
\bea
   \label{n8states}
     g^+  \, ,~~~~\lambda^+ = \lambda^{a} \, ,~~~~z^{ab}\,,~~~~
     \lambda^-=\lambda^{abc}\, , ~~~~g^- = g^{abcd}\,.
\eea
Their helicities are $h=+1,+\frac{1}{2},0,-\frac{1}{2},-1$.

The massless closed string external states are the 256 states of $\mathcal{N}=8$ supergravity, which are labeled by fully antisymmetric irreps of the $SU(8)$:
\bea
   \label{n8states}
    h^+  \, ,~~~~~\psi^+ = \psi^{a} \, ,~~~~~v^+=v^{ab}\,,
   ~~~~~\chi^+ = \chi^{abc} \, ,
~~~~\varphi^{abcd}\, ,~\dots~~~~~~~
     h^- = h^{abcdefgh}\,.
\eea
The helicities are $+2, +\frac{3}{2},\dots,-2$.
Note in particular that the scalars $\varphi^{abcd}$ sit in the {\bf 70} of $SU(8)$.\footnote{
We emphasize that $SU(8)$ is not a symmetry of the closed string tree amplitudes. It is the $R$-symmetry of $\cn=8$ supergravity, and we use it to label the massless states.}

The massless closed string states can be obtained from products of two massless open string states. We distinguish the open string states in the product by dressing the right-movers with a tilde and leaving the left-movers untilde'd. Gravitons are products of two same-helicity gluons, $h^\pm = g^\pm \otimes \tilde{g}^\pm$, half of the gravitinos come from $g^{\pm}\otimes\tilde{\l}^\pm$ while the other half come from the left/right exchanged product $\l^\pm\otimes \tilde{g}^{\pm}$. A precise map of the full spectrum can be found in Table 1 of  \cite{BEF}.

The KLT relations express the closed string tree amplitudes in terms of open string tree amplitudes. Let us here present the explicit KLT formulas for $n=4,5$:
\bea
  \begin{split}
  M_4(1,2,3,4) &\,=\, \frac{\sin(\a' \pi s_{12})}{\a' \pi} ~
  A_4(1,2,3,4)\,\widetilde{A}_4(1,2,4,3)\, ,\\[2mm]
  M_5(1,2,3,4,5)
  &\,=\,
   -\frac{ \sin(\a' \pi s_{12})   \sin(\a' \pi s_{34})}{\a'^2 \pi^2}~
   A_5(1,2,3,4,5) \,
  \widetilde{A}_5(2,1,4,3,5)
  ~+~ (2\lra 3) \, .
  \end{split}
 \eea
We use $A_n$ and $\widetilde{A}_n$  to indicate the open string amplitudes and $M_n$ for the closed string amplitudes. The KLT relations rely on the decomposition of closed string states as products of open string states that we described above.

The open string tree amplitudes respect a global $SU(4)$ symmetry. It follows directly from the decomposition of states and the KLT relations that the closed string tree amplitudes vanish unless their massless external states form an $SU(4) \times SU(4)$ invariant. In the strict field theory limit $\alpha' \to 0$, this symmetry is enhanced to the global $SU(8)$ $R$-symmetry of $\mathcal{N}=8$ supergravity. As we shall see shortly, the breaking of $SU(8)$ to $SU(4) \times SU(4)$ is manifest in the $\alpha'$-corrections to the closed string amplitude.

Finally, let us note that KLT and the decomposition of the closed string states also gives a discrete symmetry $\mathbb{Z}_2$ corresponding to the interchange of left- and right-movers; it amounts to the flip $A_n\lra \tilde{A}_n$ in the KLT relations.

%%%%%%%%%%%%
\subsection{The dilaton and its $SU(8)$-violating amplitudes}

Consider the closed string states obtained from products of gluons: two same-helicity gluons give a graviton, while combinations of opposite-helicity gluons give two scalars. One scalar is the dilaton $\phi$ and the other is the axion $a$. Specifically, the {\em canonical decomposition} $\{1,2,\dots,8\}\to \{1,2,3,4\} \otimes \{5,6,7,8\}$ gives
\bea
\label{decomp}
   h^\pm \lra g^\pm \otimes \tilde{g}^\pm \,,~~~~~~
  \varphi \equiv \varphi^{1234}\,\lra\, g^{1234} \otimes \tilde{g}^+\, ,~~~~~~
  \overline{\varphi} \equiv  \varphi^{5678}\,\lra\, g^+ \otimes \tilde{g}^{5678}\, ,
\eea
and the dilaton $\phi$ is identified as the $SU(4) \times SU(4) \times \mathbb{Z}_2$ invariant combination
\bea
  \label{phi}
  \phi = \frac{1}{2}\bigl(\varphi +  \overline{\varphi}\bigr) =  \frac{1}{2}\bigl(\varphi^{1234} +  \varphi^{5678}\bigr) \, .
\eea
As $\varphi$ and  $\overline{\varphi}$ are conjugate complex scalars, the dilaton is a real scalar field. Similarly, the axion $a=i(\varphi -  \overline{\varphi})/2$ is the real pseudo scalar associated with the orthogonal linear combination. It is easy to see that, with the normalization~(\ref{phi}), the amplitudes of $\phi$ are related to the amplitudes of $\varphi$, $\overline{\varphi}$ as
\begin{equation}
    M_n(\phi\,,\,\ldots)~=~M_n(\varphi\,,\,\ldots)\,+\,M_n(\overline{\varphi}\,,\,\ldots)\,.
\end{equation}
We can then write any dilaton amplitude in terms of amplitudes with $\varphi$ and $\overline{\varphi}$.

\vspace{2mm}
\noindent {\bf 4-point amplitudes}\\
There are no $SU(8)$-violating 4-point amplitudes. For example, amplitudes with gravitons and a single dilaton vanish because 
 they are expressed in terms of 
helicity-violating open string amplitudes,\footnote{
String amplitudes obey the same SUSY Ward identities as the field theory amplitudes \cite{STWI}.
Therefore, helicity violating superstring amplitudes $\<+\dots++\>$ and $\<+\dots+-\>$ vanish.} e.g.~
\begin{equation}
  M_4(\phi,2^+,3^+,4^-)\,=\,2\, 
  M_4(\varphi,2^+,3^+,4^-) ~\propto~
  A_4(1^-,2^+,3^+,4^-)\,\tilde{A}_4(1^+,2^+,4^-,3^+) ~=~ 0\, .
\end{equation}
Here we used the $\Z_2$ left/right exchange symmetry in the first step and KLT in the second. 
Likewise, $M_4(\varphi,\varphi,3^+,4^-)=0$. However,
amplitudes involving two dilatons have $SU(8)$ invariant contributions that are non-vanishing,
\bea
  \nonumber
  \frac{1}{2} M_4(\phi,\phi,3^+,4^-) &=&
  M_4(\varphi,\overline{\varphi},3^+,4^-)
  ~=~\frac{\sin(\a' \pi s_{12})}{\a' \pi}\, A_4(1^-,2^+,3^+,4^-)\,\tilde{A}_4(1^+,2^-,4^-,3^+)\\
  &=& \frac{\<14\>^4\<24\>^4}{\<12\>^8} M_4(1^-,2^-,3^+,4^+) \, .
\eea
The last identity follows from the fact that the amplitude
$M_4(\varphi,\overline{\varphi},3^+,4^-)$  is MHV.

\vspace{3mm}
\noindent {\bf 5-point amplitudes}\\
The simplest $SU(8)$-violating amplitude is $M_5(\varphi,2^+,3^+,4^-,5^-)$.
Schematically, KLT relates
\bea
  \label{phi5}
  M_5(\varphi,2^+,3^+,4^-,5^-)
  ~\lra~
   A_5(1^-,2^+,3^+,4^-,5^-) \,
  \widetilde{A}_5(2^+,1^+,4^-,3^+,5^-)
  ~+~ (2\lra 3)\,,
\eea
where $A_5$ is anti-MHV while $\widetilde{A}_5$ is MHV.

Since the amplitude \reef{phi5} manifestly violates $SU(8)$, it  vanishes in the field theory limit 
 $\alpha'\to 0$. 
This gives a quadratic constraint on Yang-Mills gluon amplitudes, which is recognized as one of the so-called ``vanishing identities'' recently presented in \cite{BDFS} and discussed in \cite{Feng,Tye}. Generally, the quadratic identities of  \cite{BDFS} can be identified as the field theory limit of KLT with $SU(8)$-violating gravity amplitudes. Non-trivial identities are obtained only for $SU(4) \times SU(4)$ invariant combinations of the external states.

In the $\alpha'$-expansion, the leading contribution to the amplitude \reef{phi5} enters at order $\a'^3$. The effective field theory has a corresponding local 5-point operator of dimension 8, namely $\phi\, R^4$. The 5-point matrix elements of this local operator must be polynomial in the external momenta and be of 
mass dimension 8. When expressed in spinor helicity formalism it must respect the ``little group scaling'', $M_n(\{t_i |i\>,t_i^{-1} |i],h_i\}) = t_i^{-2h_i}\,M_n(\{|i\>,|i],h_i\})$, where $h_i$ is the helicity of the $i$th particle. These properties determine the amplitude \reef{phi5} uniquely as $c\, [23]^4 \<45\>^4$. The constant $c$ can be determined using KLT. The $\a'$-expansion of the open string 5-point amplitude is readily available in the literature \cite{STMHV}, and via KLT we find
\begin{equation}
   M_5(1^+,2^+,3^-,4^-,\phi) ~=~
   2\, M_5(1^+,2^+,3^-,4^-, \varphi) 
   ~=~ 
   12\, \zeta(3) \,\a'^3\,  [12]^4 \<34\>^4 ~+~O(\a'^5)\, .
  \label{ppmmphi}
\end{equation}
There are no pole-diagram contributions to this amplitude. The entire result \reef{ppmmphi} comes from the local operator $\phi\, R^4$, whose origin is the Einstein frame operator $e^{-6 \phi} R^4$. 

\vspace{2mm}
\noindent {\bf Comment on inequivalent $ SU(4)\times SU(4)\subset SU(8)$ embeddings}\\
Before closing this section, it is worth remarking on the  $SU(4)\times SU(4)\subset SU(8)$ embeddings of the closed string spectrum. All 35 decompositions are equivalent in the field theory limit $\alpha'\to 0$. For example, we have
\be
   \label{scalarMs}
   M^{\rm sugra}_6(\varphi^{1234} \varphi^{5678}+\!+\!-\!-)\,=\,M^{\rm sugra}_6(\varphi^{12|56} \varphi^{34|78}+\!+\!-\!-)\,=\,
   -M^{\rm sugra}_6(\varphi^{123|5} \varphi^{4|678}+\!+\!-\!-) \,,~
\ee
because these three amplitudes are $SU(8)$-equivalent and therefore give identical results (up to the overall sign from odd permutations of the $SU(8)$-indices). In~(\ref{scalarMs}), and in the following, we use $|$ to separate indices that lie in distinct $SU(4)$ factors.

Relations like~(\ref{scalarMs}) do \emph{not} hold for closed string amplitudes. The dilaton, for example, selects out a preferred direction among the $70$ scalars. This special direction is reflected in the KLT relations through the choice of a left/right decomposition that selects the $SU(4)\times SU(4)$ subgroup of $SU(8)$.
In the canonical decomposition $\{1,2,\dots,8\}\to \{1,2,3,4\} \otimes \{5,6,7,8\}$, the scalars $\varphi^{1234}$ and $\varphi^{5678}$ come from products of opposite-helicity gluons \reef{decomp}, and are related to the dilaton \reef{phi}. The other scalars arise from  scalar or gluino products, specifically\footnote{The minus in the second state decomposition arises from interchanges of fermionic operators. 
}
\begin{equation}\label{2phis} 
[\varphi^{12|56} \varphi^{34|78}]  ~\, \lra~\,
  [z^{12}\, z^{34}] \otimes\, [\tilde{z}^{56}\, \tilde{z}^{78}]\,\,, \qquad \quad
-\,[\varphi^{123|5} \varphi^{4|678}]  ~\, \lra~
  \, [\l^{123}\, \l^{4}] \otimes [\tilde{\l}^{5} \,\tilde{\l}^{678}]\,.
\end{equation}
In sections  \ref{s:SSLgrp} and \ref{s:SSL} we show that the three amplitudes in \reef{scalarMs} have distinct $\a'^3$-corrections.

One could also consider non-canonical decompositions $SU(8) \to SU(4)\times SU(4)$, for example $\{1,2,\dots,8\}\to \{1,2,7,8\} \otimes \{3,4,5,6\}$. This simply rotates which scalars $\varphi$ are identified with the dilaton, etc. In this paper we will always use the canonical decomposition.

%%%%%%%%%%%%%
%%%%%%%%%%%%%
\setcounter{equation}{0}
\section{Soft scalar limits and (the elimination of) $SU(8)$-violation}
\label{s:SSLgrp}

In this section we use single-soft scalar limits to prove that the  $SU(8)$-invariant supersymmetrization of the 3-loop $R^4$ candidate counterterm of $\mathcal{N}=8$ supergravity is incompatible with $E_{7(7)}$ symmetry. MHV amplitudes have vanishing single-soft scalar limits, so we need 6-point NMHV amplitudes for non-trivial tests. Specifically, we show that the NMHV amplitude $\<\varphi\, \overline{\varphi} ++ --\>_{R^4}$ of $R^4$ is non-vanishing in the limit $p_1 \to 0$.

Our strategy is to use KLT to obtain 6-point NMHV closed string tree amplitudes with two scalars and four gravitons. At order $\a'^3$, superstring theory gives tree amplitudes of the supersymmetric operator $e^{-6\phi} R^4$. This operator has two parts: one that violates $SU(8)$ and one that preserves $SU(8)$. We systematically eliminate the $SU(8)$-violations by averaging over $SU(8)$. This leaves the desired amplitude  $\<\varphi\, \overline{\varphi} ++ --\>_{R^4}$ of the  $SU(8)$-invariant supersymmetrization of $R^4$. (In section \ref{s:SSL} we give an independent derivation based on superamplitudes.)

It is useful to introduce notation for the amplitudes studied in this section.
We write the $\a'$-expansion of the closed string amplitude as
\begin{equation}\label{norm}
    M_n(\ldots)~=~M^{\rm sugra}_n(\ldots)~-~2\zeta(3)\,\alpha'^3\bigl\<~\ldots~\bigr\>_{e^{\dash6\phi}\!R^4}~~+~O(\alpha'^5)\,.
\end{equation}
The `$\dots$' refer to the $n$ external states. In the effective action, all order $\a'^3$-corrections to the amplitude come from the $SU(4)\times SU(4)$-invariant supersymmetrization 
%he
of
 the operator $e^{-6\phi} R^4$; hence the subscript on $\<\,\ldots\,\>$.
To avoid repeating the same common factors in the $O(\a'^3)$ matrix elements, we have pulled out a factor of $-2 \zeta(3)$. With this normalization we have
\bea
  \label{4pt5pt}
  \bigl\< ++ --\bigr\>_{e^{\dash6\phi}\!R^4} ~=~ [12]^4 \<34\>^4 \, ,~~~~~~~~
  \bigl\< ++ -- \phi \bigr\>_{e^{\dash6\phi}\!R^4} ~=~ -6\times [12]^4 \<34\>^4 \, .
\eea
The first result follows from the $\a'$-expansion of the Virasoro-Shapiro-factor of the closed string $4$-graviton amplitude. The $5$-point graviton-dilaton amplitude was already presented in \reef{ppmmphi}.

%%%%%%%%%
\subsection{Warm-up: 6-point graviton-dilaton amplitude}
As a warm-up, let us start with the 6-point amplitude with two dilatons and four gravitons. In the effective field theory description, the $\alpha'^3$-terms of this amplitude receives a local contribution from the operator $\phi^2 R^4$. From the expansion of $e^{-6\phi} R^4$ and the results \reef{4pt5pt}, we immediately infer that
\bea
  \lab{M6loc}
  \bigl\< \phi \,\phi++ --  \bigr\>_{e^{\dash6\phi}\!R^4} \Big|_\text{local}
  ~=~
  36  \times [34]^4 \<56\>^4\,.
\eea
As we will see next, this local contribution accounts for the entire single-soft scalar  limits of $\bigl\<\phi \, \phi ++ --  \bigr\>_{e^{\dash6\phi}\!R^4}$.

The graviton-dilaton amplitude can be expressed in terms of amplitudes involving the scalars $\varphi$ and $\overline{\varphi}$ defined in \reef{decomp}.
Using \reef{phi} and the $\mathbb{Z}_2$-exchange symmetry, we have
\bea
  \label{6phiphi}
  M_6(\phi \,\phi++-- )
  =
  2 \, M_6(\varphi \,\varphi++-- )
  + 2\, M_6(\varphi \,\overline\varphi++-- )  \, .
\eea
The amplitude $M_6(\varphi \,\varphi++-- )$ manifestly violates $SU(8)$ and therefore vanishes in the field theory limit. Its leading contribution appears at $O(\a'^3)$. We can compute it using KLT with the state decomposition
\bea
[\varphi \,\varphi++-- ]  ~~~ \lra  ~~~
[- -++-- ] \otimes [++++--] \, .
\eea
The 6-point MHV open string amplitude is available in \cite{STMHV}, and anti-MHV is obtained by conjugation. The KLT calculation gives
\bea
  \lab{6v}
  \bigl\< \varphi \,\varphi++ --  \bigr\>_{e^{\dash6\phi}\!R^4}
  ~=~ 12 \times [34]^4 \<56\>^4 \, .
\eea
Locality of this result is consistent with the Feynman rules of $e^{-6\phi} R^4$; no pole terms can contribute at this order.

The external states of the last amplitude  in \reef{6phiphi}, $M_6(\varphi \,\overline\varphi++--)$, 
 include an $SU(8)$-singlet, and it is therefore non-vanishing for $\a'=0$. The supergravity contribution can be calculated from recursion relations, and we checked that it matches the $\alpha'\to0$ limit of the amplitude obtained from KLT. The needed KLT state decomposition is
\bea
[\varphi \,\overline{\varphi}++-- ]  ~~~ \lra  ~~~
[- +++-- ] \otimes [+-++--]\, .
\eea
The NMHV open string gluon amplitudes can be found in \cite{ST6pt}.
On the closed string side, the leading $\a'$-correction is at order $\a'^3$. From the point of view of the effective field theory, it contains non-local terms arising from pole diagrams with 4- and 5-point vertices from $e^{-6\phi} R^4$, as well as local 6-point contact terms.
We find  that the single-soft scalar limit $p_1\to0$ is particularly simple:
\bea
  \label{softp}
   \lim_{p_1\to0} \,
   \bigl\< \varphi \,\overline\varphi++ --  \bigr\>_{e^{\dash6\phi}\!R^4}
   ~=~ 6\times [34]^4 \<56\>^4.
\eea
Now combine \reef{6phiphi}, \reef{6v} and \reef{softp} to find
\bea
  \nonumber
  \lim_{p_1\to0} \,\bigl\< \phi \,\phi++ -- \bigr\>_{e^{\dash6\phi}\!R^4}
  &=&
  2 \, \lim_{p_1\to0} \,\bigl\< \varphi \,\varphi++ --  \bigr\>_{e^{\dash6\phi}\!R^4}
  + 2\, \lim_{p_1\to0} \,\bigl\< \varphi \,\overline\varphi++ --  \bigr\>_{e^{\dash6\phi}\!R^4} \\
  \nonumber
   &=& 36\,\times [34]^4 \<56\>^4 \, .
\eea
We compare with \reef{M6loc} and note that the entire contribution to the non-vanishing single-soft scalar limit arises from the local operator $\phi^2 R^4$. One should of course not be surprised that the dilaton amplitude has non-vanishing single-soft scalar limits: it is associated with the local operator $e^{-6\phi}R^4$, which violates $SU(8)$, so it is not compatible with $E_{7(7)}$.

The point of this subsection was to illustrate the role of the dilaton in the non-vanishing single-soft scalar limit of the closed string amplitude $M_6(++ -- \varphi \,\overline\varphi)$. This amplitude contains both $SU(8)$-preserving and $SU(8)$-violating contributions. Our task in the next section is to isolate the $SU(8)$-singlet contribution.

%%%%%%%%
\subsection{Construction of  $SU(8)$-invariant matrix elements and soft scalar limits}

It is convenient to reintroduce the $SU(8)$ indices of $\varphi=\varphi^{1234}$ and
$\overline{\varphi}=\varphi^{5678}$. The decomposition of the scalar polynomial $\varphi^{1234} \varphi^{5678}$ contains an $SU(8)$-singlet as well as components that are only $SU(4) \times SU(4)$-invariant. The $SU(8)$-singlet is
\bea
   \label{singlet}
   (\varphi \overline{\varphi})_\text{singlet} &=&\frac{1}{8!}\,\eps_{a_1a_2a_3a_4 a_5a_6a_7a_8} \,\varphi^{a_1a_2a_3a_4} \,\varphi^{ a_5a_6a_7a_8}
   ~=~\frac{1}{35}  \,\,\text{Sym}_{35}(\varphi^{1234} \varphi^{5678})
    \, ,
\eea
where Sym$_{35}$ stands for the sum over the $35$ inequivalent polynomials obtained from antisymmetrization of the indices $\{1,2,\dots,8\}$.

The $SU(8)$-invariant two-scalar amplitude $\<\varphi\, \overline{\varphi} ++ -- \>$ of $R^4$ is given by the singlet contribution to the corresponding two-scalar amplitude of $e^{-6\phi}R^4$:
\begin{equation}
  \big\<\varphi\, \overline{\varphi} ++ -- \big\>_{R^4}
  ~\equiv~
  \big\< (\varphi \overline{\varphi})_\text{singlet} ++ -- \big\>_{e^{\dash6\phi}\!R^4} \,.
\end{equation}
Averaging the matrix elements with two scalars and four gravitons over the 35 distinct arrangements of  $SU(8)$-indices given in~(\ref{singlet}),  we get
\bea
\nonumber
  \big\<\varphi\, \overline{\varphi} ++ -- \big\>_{R^4}
  \!&=&\!
\nonumber
   \frac{1}{35}\big\< \varphi^{1234} \varphi^{5678} ++ --\big\>_{e^{\dash6\phi}\!R^4}
   \,-\, \frac{16}{35} \, \big\<\varphi^{123|5} \varphi^{4|678} ++ --\big\>_{e^{\dash6\phi}\!R^4}
   \\[2mm]
   &&
   ~~~\,+\,\frac{18}{35}\, \big\<\varphi^{12|56} \varphi^{34|78} ++ --\big\>_{e^{\dash6\phi}\!R^4}\,.
  \label{rel2}
\eea
We have exploited $SU(4) \times SU(4)$-invariance of the closed string tree amplitude to organize the matrix elements into three groups with 
 inequivalent 
$SU(4) \times SU(4)$ structures.

All we have to do now is swing the three amplitudes on the right-hand side of~(\ref{rel2}) through KLT; then we have the desired singlet contribution and we can calculate its soft scalar limit.
The decompositions of the closed string scalars of \reef{rel2} into open string states were given in \reef{decomp} and \reef{2phis}. The required open string amplitudes 
are all in the NMHV sector. They are not all among the 12 open string amplitudes presented by Stieberger and Taylor \cite{ST6pt}. We use the solution to the $\mathcal{N}=4$ SYM SUSY Ward identities \cite{efk4} to express the 5 basis amplitudes in eq.~(3.13) of \cite{efk4} in terms of 5 amplitudes from \cite{ST6pt}. This, together with KLT, then allows us to automate (in Mathematica) the process of obtaining any  open or closed  string 6-point NMHV tree amplitude.\footnote{The fact that the solutions to the SUSY Ward identities for 6-point amplitudes only require 5 independent basis amplitudes 
 allowed 
us to check 7 non-trivial relationships among the 12 open string amplitudes of \cite{ST6pt}.}
For the three scalar-graviton amplitudes \reef{rel2}, the result is a set of complicated amplitudes at $O(\a'^3)$, but their single-soft scalar limits are very simple:\footnote{We determine these limits numerically and find convergence towards the results \reef{softp3} as the momentum $p_1$ is decreased. Working at high numerical precision we come within $10^{-40}$ of the shown integer coefficients for several different values of the external momenta.}
\bea
  \label{softp3}
   \lim_{p_1\to0} \, \big\< \varphi^{1234} \varphi^{5678} ++ --\big\>_{e^{\dash6\phi}\!R^4}
     &=&6\times [34]^4 \<56\>^4,\nonumber\\
   \lim_{p_1\to0} \, \big\<\varphi^{12|56} \varphi^{34|78} ++ --\big\>_{e^{\dash6\phi}\!R^4}
     &=&  0\,,\\
   \lim_{p_1\to0} \, \big\<\varphi^{123|5} \varphi^{4|678} ++ --\big\>_{e^{\dash6\phi}\!R^4}
     &=& 3\times   [34]^4 \<56\>^4\,.\nonumber
\eea
We can now combine the results  \reef{rel2} and \reef{softp3}
to obtain the single-soft scalar limits of $R^4$:
\be\label{soft2}
\boxed{\phantom{\Biggl{(}}
    \lim_{p_1\to0} \, \big\<\varphi\, \overline{\varphi} ++ -- \big\>_{R^4} ~=~
   -\frac{6}{5}\, [34]^4 \<56\>^4
   ~\ne~ 0\,.
   ~~}
\ee
Since the matrix elements of $R^4$ are $SU(8)$-invariant, \reef{soft2} holds for any of the 35 pairs of conjugate scalars.
The single-soft scalar limits of $R^4$ are non-vanishing, and this demonstrates the incompatibility of the $SU(8)$ supersymmetrization of $R^4$ with $E_{7(7)}$-symmetry. It is therefore excluded as a candidate counterterm in perturbative $\mathcal{N}=8$ supergravity.

%%%%%%%%%%
\subsection{Matching of soft scalar limits to automorphism analysis}

The moduli dependence of operators of the form $f(\varphi)D^{2k}R^4$ 
 was 
studied in \cite{Eisenstein,Pioline}. In four dimensions, $f(\varphi)$ must satisfy the Laplace equation
\begin{equation}\label{LapEq}
   \Delta f(\varphi)~=~\lambda  f(\varphi)\,,\quad\qquad\text{with }\quad \lambda=\frac{3(11-D)(D-8)}{D-2}\Biggr|_{D=4}\!\!=\,-42\,.
\end{equation}
$\Delta$ is the Laplacian on the coset space $E_{7(7)}/SU(8)$, which is parameterized by the scalars $\varphi^{ijkl}$. In our normalization of the 70 scalars $\varphi^{ijkl}$, the Laplacian takes the form\footnote{
For the complex scalar field $\Omega$ of a non-linear sigma model with action $(1/\ell_P^{2})\int\!\sqrt{-g}\, g_{\Omega \bar \Omega}\,\partial_\m \Omega\, \partial^\m \bar \Omega$\,,
the Laplacian in~\cite{Eisenstein} is normalized as $\Delta=2 g^{\bar\Omega \Omega} \partial_{\bar\Omega}\partial_{\Omega}+\,\ldots$\,. In our normalization, the four-dimensional Planck length is set to $\ell_P=\sqrt{2}$ to avoid factors $(\sqrt{2}/\ell_P)^{n-2}$ in KLT.  Also, all 35 pairs of conjugate
scalars $\varphi$, $\bar \varphi$  are canonically normalized as $\int\!\sqrt{-g}\,\partial_\mu \varphi \partial^\mu \bar \varphi +\dots$\,. The normalizations for complex scalar fields are thus related by $\Omega\to \sqrt{2}\varphi$. This implies that the Laplacian $\Delta$ of~\cite{Eisenstein} takes the form~(\ref{Laplace}) when expressed in terms of the scalars $\varphi^{ijkl}$.}
\begin{equation}\label{Laplace}
    \Delta~=~\Big( \frac{\partial}{\partial \varphi^{1234}}\frac{\partial}{\partial \varphi^{5678}}+\text{34 inequivalent perms} \Big)~~+~\ldots\,
\end{equation}
to leading-order, i.e.~the terms `$\ldots$' contain higher powers of the fields $\varphi$. Our $6$-point analysis is not sensitive to these corrections.

The results
\bea
 \bigl\< ++ --\bigr\>_{R^4} ~=~ [12]^4 \<34\>^4
 ~~~~\text{and}~~~~~
  \lim_{p_1\to0} \, \big\<\varphi\, \overline{\varphi} ++ -- \big\>_{R^4} ~=~
   -\frac{6}{5}\, [34]^4 \<56\>^4 \,
\eea
determine the $SU(8)$-invariant moduli-function $f_{R^4}(\varphi)$ of $R^4$ to quadratic order in the scalars:
\begin{equation}
    f_{R^4}(\varphi)~=~1-\frac{6}{5}\,\Bigl(\varphi^{1234} \varphi^{5678}+\text{34 inequivalent perms}\Bigr)+O(\varphi^4)\,.
\end{equation}
It follows that
\begin{equation}
    \Delta f_{R^4}(\varphi) ~=~ -\frac{6}{5}\times 35+O(\varphi^2)~=~-42\times\Bigl(1 +O(\varphi^2)\Bigr)\,.
\end{equation}
The right-hand side is $\l \,  f_{R^4}$ to leading order. Thus we find that the non-linear $SU(8)$-invariant completion of the supersymmetric operator $R^4$ indeed satisfies the constraint~(\ref{LapEq}) up to the order that we constructed it.\footnote{We have also performed this match at order $\a'^5$ \cite{future}, 
 which provides another independent check on the normalization of the Laplacian~(\ref{Laplace}).
}
%he in footnote I have updated ref.

%%%%%%%%%%%%%
\setcounter{equation}{0}
\section{Superamplitudes}
\label{s:superA}

\subsection{N$^{(k,\tilde{k})}$MHV classification and super-KLT}
A generalization of the familiar N$^k$MHV classification is required to accommodate closed string processes that violate $SU(8)$. Let us first note that in supersymmetric theories without $R$-symmetry, amplitudes can in principle have a different  N$^{k_a}$MHV level for each supercharge $Q^a$,  $a=1,\ldots,\cn$.\footnote{It is a consequence of supersymmetry that amplitudes of massless particles vanish unless $0\leq k_a\leq n- 4$.}
In supergravity, the $SU(8)$ $R$-symmetry rotates the supercharges and hence all amplitudes are classified by a single integer $k$.  It is useful to encode 
 supergravity 
amplitudes in the same N$^k$MHV sector into superamplitudes, which are Grassmann polynomials of degree $8+8k$.

The closed string tree amplitudes respect $SU(4) \times SU(4)$ $R$-symmetry.  The supercharges $Q^1,\ldots,Q^4$ can be rotated into each other and therefore have the same N$^k$MHV level. Similarly, the supercharges $Q^5,\ldots,Q^8$ have the same N$^{\tk}$MHV level. Hence two integers  $k$ and $\tilde{k}$ are needed to  characterize the amplitudes; we denote the sectors by N$^{(k,\tilde{k})}$MHV. The corresponding superamplitudes are Grassmann polynomials of degree $8+4k+4\tilde{k}$.

$SU(8)$-invariant N$^k$MHV amplitudes vanish unless the external states, labeled with upper indices only, arrange themselves into an integer number of 8-index Levi-Civita tensors $(\eps^{........})^{k+2}$. For the case of $SU(4) \times SU(4)$-invariance, the tensor structure  $(\eps^{....})^{k+2}(\tilde{\eps}^{....})^{\tk+2}$ characterizes the N$^{(k,\tilde{k})}$MHV sector.

In the KLT relations, N$^{(k,\tilde{k})}$MHV closed string tree amplitudes are products of open string amplitudes that lie in the N$^{k}$MHV and N$^{\tilde{k}}$MHV sectors, respectively.
KLT is then naturally promoted to a super-KLT relation relating superamplitudes:\footnote{This type of KLT decomposition was also recently used in~\cite{Feng}.}
\bea\label{superKLT} 
    \sum_{k,\tk} \mathcal{M}_n^{(k,\tk)}
    = \sum_\mathcal{P} f_n(s_{ij}) ~
       \Big( \sum_{k} \mathcal{A}_n^{k}\Big) \times
       \Big(\sum_{\tk} \tilde{\mathcal{A}}_n^{\tk}\Big) \,.
\eea
In $\cn=8$ supergravity we are familiar with N$^k$MHV=N$^{(k,k)}$MHV amplitudes. Amplitudes with $k \ne \tk$ necessarily break $SU(8)$ and therefore vanish when $\a'\to0$.
We have seen explicit examples in sections \ref{2} and \ref{s:SSLgrp}. For example, the amplitude $M_5(\varphi ++--)$ is N$^{(1,0)}$MHV, $M_6(\varphi \varphi ++--)$ is N$^{(2,0)}$MHV, and $M_6(\varphi \overline{\varphi} ++--)$  is N$^{(1,1)}$MHV = NMHV.
In the field theory limit, the vanishing of the $SU(8)$-violating cross-terms with $k \ne \tk$ yields a superamplitude version of the ``vanishing identities'' mentioned earlier. We discuss 
 stringy generalizations 
of these 
 identities 
in section \ref{s:vanishing}.

Amplitudes in supersymmetric theories are related through the Ward identities of supersymmetry and $R$-symmetry. For $\cn=4$ SYM and $\cn=8$ supergravity, these constraints have been solved in full generality in~\cite{efk4}. This gives a manifestly supersymmetric and $SU(\cn)$-symmetric expression for the superamplitudes in terms of a set of independent basis amplitudes. Now we present a similar solution to the Ward identities for the simplest cases of superamplitudes whose $R$-symmetry is reduced to $SU(4)\times SU(4)$.

\subsection{MHV, $\sqNMHV$, and N'MHV superamplitudes}
It is well-known that the MHV superamplitudes of $\cn=8$ supergravity take the form
\begin{equation}\label{MHVsup}
    \mathcal{M}_n^{\rm MHV}~=~\frac{\delta^{(16)}\bigl(\sum_i|i\>\eta_{ia}\bigr)}{\<n-1,n\>^8}\,\times\,M_n(+\ldots +--)\,.
\end{equation}
The $\eta_{ia}$'s are Grassmann variables\footnote{For details of the notation and SUSY Ward identities we refer the reader to \cite{efk4}.} labeled by momenta $i=1,\dots,n$  and 
 $SU(8)$ indices 
$a=1,\dots,8$. It follows from \reef{MHVsup} that the MHV sector is completely determined by the single basis amplitude $M_n(+\ldots +--)$.
 For closed string amplitudes in the MHV (\ie N$^{(0,0)}$MHV) sector, the solution 
  to the Ward identities is also just 
\reef{MHVsup}: the $\eta$-dependence 
 of 
each of the open string MHV 
 superamplitudes is in 
the Grassmann delta-function, so KLT simply gives
$\delta^{(8)}\times \tilde{\delta}^{(8)}=\delta^{(16)}$.

The first true $SU(4)\times SU(4)$ superamplitude sits at the N$^{(1,0)}$MHV$\oplus$N$^{(0,1)}$MHV level, which we call the 
 ``{\bf$\sqrt{\text{{\bf N}}}$MHV sector}'', 
for simplicity. As we have seen, it first occurs at the $5$-point level, where it can be understood as a product of MHV and anti-MHV open string amplitudes.
Clearly, $\sqNMHV$ manifestly breaks $SU(8)$. Our construction of the $\sqNMHV$ superamplitude proceeds very similarly to the construction in~\cite{efk4} of NMHV superamplitudes in $\cn=4$ SYM.
Defining the $Q$-invariant polynomials
\begin{equation}
    m_{ijk,a} \equiv [ij] \eta_{ka} +[jk] \eta_{ia} +[ki] \eta_{ja}\,,
\end{equation}
we can construct
\begin{equation}\label{introdefYs}
   Y_{ijkl} ~=~
 [n\!-\!3,n\!-\! 2]^{-4} 
 \times \,m_{i,n\dash 3,n\dash 2;1}\,\,m_{j,n\dash 3,n\dash2;2}\,\,m_{k,n\dash 3,n\dash 2;3}\,m_{l,n\dash 3,n\dash 2;4}\,\,\,.
\end{equation}
Clearly, $Y_{ijkl}$ is invariant under the $SU(4)$ transformations acting on $a=5,6,7,8$, because it does not depend on 
 the $\eta_{ia}$ of 
this sector. After full symmetrization of its indices,  $Y_{(ijkl)}$ is even a fully $SU(4)\times SU(4)$ invariant polynomial. In the following, we will also need  the completely analogous polynomial $\tilde{Y}_{ijkl}$ that 
 depends on the $\eta_{ia}$ in the other $SU(4)$ sector and 
arises from $Y_{ijkl}$ under the $\Z_2$ flip.

The $n$-point  $\sqNMHV$ superamplitude then takes the form
\begin{equation}\label{sqNMHV}
    {\cal M}_n^\sqNMHV~=~\frac{\delta^{(16)}\bigl(\sum_i | i\> \eta_{ai}\bigr)}{\<n-1,n\>^8 }\times\!\!\!
    \sum_{1\leq i\leq j\leq k\leq l\leq n-4}\!\!\!M_n(\{i,j,k,l\}++--) \,\,\Bigl[Y_{(ijkl)}+
    \tilde{Y}_{(ijkl)}\Bigr]\,, 
\end{equation}
where the $\Z_2$ symmetry ensured that $Y$ and $\tilde{Y}$ are multiplied by the same basis amplitudes.
The basis amplitudes $M_n(\{i,j,k,l\}++--)$ have the indicated gravitons on the last four lines. Their particle content on the remaining lines is determined by the set $\{i,j,k,l\}$, which
indicates that state $i$ carries $SU(4)$ index 1, state $j$ carries $SU(4)$ index 2, and so on.  Let us illustrate this with two examples:
\begin{equation}
    M_6(\{1,1,1,1\}++--)\,\equiv\,M_6(\varphi^{1234}+++--)\,, \quad M_n(\{1,1,2,2\}++--)\,\equiv\,M_n(v^{12}v^{34}++--)\,.
\end{equation}

In general, there are then ${n-1 \choose 4}$ basis amplitudes at the $\sqNMHV$ level.
At the 5-point level, for example, there is precisely  one  basis amplitude, namely $M_5(\{1,1,1,1\}++--)=M_5(\varphi^{1234}++--)$, and the superamplitude
is given by
\begin{equation}
\begin{split}
    {\cal M}_5^\sqNMHV
    &=\frac{\delta^{(16)}\bigl(\sum_i | i\> \eta_{ai}\bigr)}{\<45\>^8}\times M_5(\varphi^{1234}++--)\bigl[Y_{1111}+\tilde{Y}_{1111}\bigr]\,.
\end{split}
\end{equation}

Next we discuss the N$^{(2,0)}$MHV$\oplus$N$^{(0,2)}$MHV sector, which we denote 
{\bf N'MHV}  
for short. Let us consider only the $n=6$ case. Then this sector comes from anti-MHV$\times$MHV in super-KLT. The superamplitude requires only one basis amplitude, $M_6(\varphi^{1234}\varphi^{1234}++--)$. The superamplitude is given by
\begin{equation}\label{NpMHV}
    {\cal M}_6^{\rm N'MHV}~=~ \frac{\delta^{(16)}\bigl(\sum_i | i\> \eta_{ai}\bigr)}{\<n-1,n\>^8}
    \times M_6(\varphi^{1234}\varphi^{1234}++--)\Bigl[Y_{1111}Y_{2222}+\tilde{Y}_{1111}\tilde{Y}_{2222}\Bigr]\,. 
\end{equation}
This superamplitude manifestly violates $SU(8)$ and thus vanishes in the supergravity limit.

%%%%%%%%%%%%%
\subsection{The NMHV sector}

Superamplitudes in the N$^{(1,1)}$MHV (=NMHV) sector are even non-vanishing when the full $SU(8)$ symmetry is imposed, so their analysis is more subtle. The breaking to $SU(4)\times SU(4)$, however, allows amplitudes to be distinct even if their particle 
content  
is related by $SU(8)$. This was illustrated with the example of three scalar amplitudes in \reef{scalarMs}.
 Thus, more basis amplitudes are needed to fully specify NMHV superamplitudes with reduced $R$-symmetry.
 The general NMHV superamplitude takes the form
\begin{equation}\label{CsqNMHV}
    {\cal M}_n^\text{NMHV}~=~\frac{\delta^{(16)}\bigl(\sum_i | i\> \eta_{ai}\bigr)}{\<n-1,n\>^8 }\!\!\!
    \sum_{{}^{1\leq i\leq j\leq k\leq l\leq n-4}_{1\leq p\leq q\leq u\leq v \leq n-4}}\!\!\!M_n({\{i,j,k,l|p,q,u,v\}};++--) \,Y_{(ijkl)}\tilde{Y}_{(pquv)}\,,
\end{equation}
where the set $\{i,j,k,l|p,q,u,v\}$ represents the lines on which the remaining indices $a=1,2,\ldots,8$ are distributed. For example,
\begin{equation}
    M_6({\{1,1,2,2|1,2,2,2\}};++--)~=~M_6(\chi^{12|5}\chi^{34|678}++--)\,.
\end{equation}
If, in addition, the $\Z_2$ symmetry is imposed, we have
\begin{equation}
    M_n({\{i,j,k,l|p,q,u,v\}};++--)~=~M_n({\{p,q,u,v|i,j,k,l\}};++--)\,.
\end{equation}
At 6-point, with $\Z_2$ symmetry, we then have 15 instead of 9 algebraic basis amplitudes. These are still not {\em functionally} independent, because we can exchange lines $1$ and $2$ due to the permutation symmetry in these lines. The functionally independent basis of matrix elements with $SU(4)\times SU(4)\times \Z_2$ symmetry then consists of only $9$ amplitudes, which are given by
\begin{equation}\label{bs}
\begin{split}
    &M_6(+-++--)\,, ~  \qquad \qquad \quad M_6(\psi^1\psi^{234|5678}++--)\,,\qquad~ M_6(v^{12}v^{34|5678}++--)\,,\\
    &M_6(v^{1|5}v^{234|678}++--)\,,\qquad  \,M_6(\chi^{123}\chi^{4|5678}++--)\,, \qquad\!\! M_6(\chi^{12|5}\chi^{34|678}++--)\,,\\
    &M_6(\varphi^{1234}\varphi^{5678}++--)\,,\qquad \, M_6(\varphi^{123|5}\varphi^{4|678}++--)\,, \qquad ~M_6(\varphi^{12|56}\varphi^{34|78}++--)\,.
\end{split}
\end{equation}
Imposing full $SU(8)$ is equivalent to demanding
\begin{equation}
\begin{split}\label{SU8cond}
     M_6(v^{12}v^{34|5678}++--)&=-M_6(v^{1|5}v^{234|678}++--)\,,\\
    M_6(\chi^{123}\chi^{4|5678}++--)&=M_6(\chi^{12|5}\chi^{34|678}++--)\,, \\
    M_6(\varphi^{1234}\varphi^{5678}++--)&=M_6(\varphi^{12|56}\varphi^{34|78}++--)=-M_6(\varphi^{123|5}\varphi^{4|678}++--)\,.
\end{split}
\end{equation}
These $SU(8)$ conditions reduce the number of functionally independent basis elements to 5, as expected from the results of~\cite{efk4}.

%%%%%%%%%%%%%%%%%%%%%%%%
\setcounter{equation}{0}
\section{Local SUSY operators with $SU(4)\times SU(4)$ symmetry}\label{s:local}

The superamplitude formulation of $SU(4)\times SU(4)$-symmetric amplitudes can also be used to encode the matrix elements of local operators. In~\cite{efk5}, superamplitudes for the matrix elements of local supersymmetric operators were constructed systematically for the case of unbroken $R$-symmetry. The basic ingredient of the construction \cite{efk5} is the fact that local on-shell matrix elements can be expressed as polynomials in angle- and square brackets $\<ij\>$ and $[ij]$. Little-group transformation properties and the dimension of the operator then constrain these polynomials, and a general ansatz can be written down for each basis amplitude. Permutation symmetry is imposed on the resulting ansatz of the superamplitude, and as shown in \cite{efk5} this guarantees that all matrix elements it produces are local.

It is known from various approaches~\cite{Harmonic,efk5} that there is a unique $SU(8)$-invariant local operator of mass dimension 8, namely the supersymmetrization of $R^4$. In particular, there are no \emph{independent} $SU(8)$-invariant local SUSY operators composed of $5$- and $6$ fields.
{\bf MHV} amplitudes are automatically $SU(8)$ invariant. We now construct superamplitudes of local operators with $SU(4)\times SU(4)$ symmetry in the simplest sectors beyond the MHV level.

\noindent {\bf $\sqrt{\text{N}}$MHV:} There exists a unique independent $5$-point operator of dimension 8 with reduced $SU(4)\times SU(4)\times \Z_2$ $R$-symmetry. It gives local 5-point amplitudes in the $\sqrt{\text{N}}$MHV sector. Not surprisingly, this operator corresponds to the supersymmetrization of the operator $\phi R^4$
 in 
the closed string effective action. We know from~(\ref{sqNMHV}) that the superamplitude is fully specified by a single basis matrix element $M_5$, which is given by
\begin{equation}
    M_5(\varphi^{1234}++--)=[23]^4\<45\>^4\,.
\end{equation}
Thus we obtain the following superamplitude
\begin{equation}\label{O5}
    {\cal O}_5^\sqNMHV
    =\delta^{(16)}\Bigl(\sum_i | i\> \eta_{ai}\Bigr)
    \times [23]^4\<45\>^{-4}\bigl[Y_{(1111)}+\tilde{Y}_{(1111)}\bigr]\,,
\end{equation}
which encodes the local contributions of the independent supersymmetrization of $\phi R^4$. We used the methods of~\cite{efk5} to verify that~(\ref{O5}) is a local and permutation-symmetric superamplitude.

At the 6-point level, there are two  
 independent 
local SUSY operators 
 with $SU(4)\times SU(4)$ symmetry.  
One operator contributes only at the N'MHV level and the other only at the NMHV level. The latter will be of particular interest to us in the next section.

\noindent {\bf N'MHV:} We find that there is a unique local SUSY operator of dimension 8 in the N'MHV sector. Its basis amplitude is
\begin{equation}\label{ONpMHVbasis}
    M_6(\varphi^{1234}\varphi^{1234}++--)~=~[34]^4\<56\>^4,
\end{equation}
and thus
\begin{equation}\label{ONpMHVsuper}
    {\cal O}_6^{\rm N'MHV}~=~\delta^{(16)}\Bigl(\sum_i | i\> \eta_{ai}\Bigr)
    \times [34]^4\<56\>^{-4}\Bigl[Y_{1111}Y_{2222}+
%he fixed indices Y_{22222} above and below
    \tilde{Y}_{1111}\tilde{Y}_{2222}\Bigr]\,.
\end{equation}
Up to an overall normalization, this operator represents the N'MHV contribution of $\phi^2R^4$.

\noindent {\bf NMHV:}
The second independent $6$-point operator of dimension $8$ with $SU(4)\times SU(4)\times\Z_2$ symmetry is in the NMHV sector.
Nine basis matrix elements \reef{bs} are needed to determine 
 its $6$-point 
NMHV superamplitude. However, for our operator here only the three scalar basis matrix elements are non-vanishing. Their momentum dependence is uniquely fixed by the mass dimension and the little group scaling. Permutation symmetry of the superamplitude fixes their relative coefficients. We find
\begin{equation} 
\begin{split}\label{O6basis}
   M_6(\,\varphi^{1234}\,\varphi^{5678}\,++--)~=~&12\times[34]^4\<56\>^4\,,\\
   M_6(\varphi^{12|56}\varphi^{34|78}++--)~=~&~\,2\times[34]^4\<56\>^4\,,\\
   M_6(\varphi^{123|5}\varphi^{4|678}++--)~=~&~\,3\times[34]^4\<56\>^4\,.
\end{split}
\end{equation}
Since these scalar amplitudes have different numerical coefficients, the $SU(8)$ violation of this operator is manifest. The explicit superamplitude then reads
\begin{equation}
\begin{split}\label{O6NMHVsuper}
        {\cal O}_6^\text{NMHV}
        ~=~\delta^{(16)}\bigl(\sum_i | i\> \eta_{ai}\bigr)
        \times [34]^4\<56\>^{-4} \Bigl[&12\bigl(Y_{1111}\tilde{Y}_{2222}+Y_{2222}\tilde{Y}_{1111}\bigr)
        +2\,Y_{(1122)}\tilde{Y}_{(1122)}\\ 
        &-3\bigl(Y_{(1112)}\tilde{Y}_{(1222)}+Y_{(1222)}\tilde{Y}_{(1112)}\bigr)\Bigr]\,.
\end{split}
\end{equation}

The $SU(4)\times SU(4)\times\Z_2$-invariant superamplitude~(\ref{O6NMHVsuper}) encodes  the $SU(8)$-{\em violating} component of $\phi^2R^4$ in the NMHV sector.
The $SU(8)$-{\em preserving} component of $\phi^2R^4$ is automatically part of the supersymmetrization of $R^4$, and therefore does not  give rise to another 
 {\em independent} local  
6-point SUSY operator.\footnote{At dimension 8, setting all three scalar basis elements equal to each other gives a superamplitude that produces non-local matrix elements.
%attempted to set all matrix elements in~(\ref{O6basis}) equal to each other, the resulting superamplitude would produce non-local matrix elements.
}  One can also check directly that no $SU(8)$-singlet contributions `hide' in \reef{O6NMHVsuper}: if we average the superamplitude~(\ref{O6NMHVsuper}) over $SU(8)$, we find that it is  $\propto12+2\times 18 - 3 \times 16 = 0$.

\setcounter{equation}{0}
\section{$e^{-6\phi}R^4$ versus $R^4$ at the $6$-point level}
\label{s:SSL}

In this section we derive the 6-point NMHV superamplitude associated with the $SU(8)$-invariant supersymmetrization of $R^4$. This is done by systematically subtracting $SU(8)$-violating contributions from the matrix elements of $e^{-6\phi}R^4$ arising from the $\a'$-expansion of the closed string tree amplitude.  As $R^4$ is the unique dimension-8 operator with unbroken $SU(8)$ symmetry~\cite{Harmonic,efk5}, this procedure is guaranteed to yield the matrix elements of $R^4$. The analysis here gives an independent derivation of the non-vanishing soft-scalar limits found in section \ref{s:SSLgrp}.

We use the convention \reef{norm} for the  matrix elements $\bigl\<~\ldots~\bigr\>_{e^{\dash6\phi}\!R^4}$ of $e^{-6\phi}R^4$. The matrix elements of the $SU(8)$-invariant supersymmetrization of $R^4$ are denoted by $\bigl\<~\ldots~\bigr\>_{R^4}$ and they are normalized by the {\bf MHV} sector
\begin{equation}
    \bigl\<~\ldots~\bigr\>_{\!R^4}^{\rm MHV} ~\equiv~\bigl\<~\ldots~\bigr\>_{e^{\dash6\phi}\!R^4}^{\rm MHV}\,,
\end{equation}
where $SU(8)$ is not violated.
An explicit check shows that all single-soft scalar limits vanish at the 4-, 5- and 6-point MHV level. As a warm-up for NMHV, we briefly discuss the $\sqNMHV$ and N'MHV sectors.

\noindent {\bf $\sqrt{\text{N}}$MHV}:
The 6-point $\sqNMHV$ matrix elements of $e^{-6\phi}R^4$ are non-vanishing because they receive non-local contributions from Feynman diagrams involving a $\phi R^4$ vertex and a supergravity $3$-point vertex. Since the $\sqNMHV$ sector manifestly violates $SU(8)$, ``subtracting the $SU(8)$-violation from $e^{-6\phi}R^4$'' is trivial in this case:
 the 
$R^4$ matrix elements in this sector simply vanish, 
$\bigl\<\,\ldots\,\bigr\>_{\!R^4}^{\sqNMHV} \,\equiv\,0$.

\noindent {\bf N'MHV}: This sector manifestly violates $SU(8)$, so $R^4$ cannot have any such contributions, and hence $\bigl\<\,\ldots\,\bigr\>_{\!R^4}^{\rm N'MHV} \,\equiv\,0$.
However, it is instructive to identify the non-vanishing contribution of $e^{-6\phi}R^4$ in this sector. Non-local terms involving a $\phi R^4$-vertex cannot contribute at order $\alpha'^3$ in this sector, and the only contributions to the N'MHV matrix elements stem from $\phi^2R^4$. These 6-point matrix elements are of course local.  Up to overall normalization, the 6-point N'MHV superamplitude is equal to that of the unique local $SU(4)\times SU(4)$ SUSY operator in the $6$-point N'MHV sector, ${\cal O}_6^{\rm N'MHV}$, which we identified in section~\ref{s:superA}; see \reef{NpMHV}.  
We used KLT to calculate the single required basis amplitude \reef{6v}, and find that  the $6$-point N'MHV superamplitude of $e^{-6\phi}R^4$ is given by 
\begin{equation}
    {\cal M}_{6,\,e^{\dash6\phi}\!R^4}^{\rm N'MHV}~=~12\times{\cal O}_6^{\rm N'\!MHV}\,.
\end{equation}
It is obvious from~(\ref{6v}) that amplitudes from $e^{-6\phi}R^4$ have non-vanishing single-soft scalar limits, but the entire N'MHV sector is absent for $R^4$, so this does not affect $R^4$.

%%%%%%%%
\subsection{The NMHV sector}
It takes 9 basis matrix elements \reef{bs} to determine the  NMHV 6-point superamplitude \reef{CsqNMHV} with $SU(4) \times SU(4)$ symmetry. To use string amplitudes to construct these basis matrix elements of $R^4$, we first need to identify the $SU(8)$-violation of $e^{-6\phi}R^4$. $SU(8)$ requires the relations \reef{SU8cond} to hold among the basis amplitudes of $R^4$. An explicit KLT computation shows that
 two of these relations, 
\begin{equation}
\begin{split}\label{nscffR4}
    \bigl\<v^{12}v^{34|5678}++--\bigr\>_{e^{\dash6\phi}\!R^4}~&=\,-\bigl\<v^{1|5}v^{234|678}++--\bigr\>_{e^{\dash6\phi}\!R^4}\,,\\
    \bigl\<\chi^{123}\chi^{4|5678}++--\bigr\>_{e^{\dash6\phi}\!R^4}~&=~\,\,\bigl\<\chi^{12|5}\chi^{34|678}++--\bigr\>_{e^{\dash6\phi}\!R^4}\,,
\end{split}
\end{equation}
 are satisfied by the matrix elements of $e^{-6\phi} R^4$. 
This means that the only $SU(8)$-violation occurs in the 3 scalar basis amplitudes; indeed  KLT helps us to show that
\begin{equation}\label{scffR4}
    \bigl\<\varphi^{1234}\varphi^{5678}++--\bigr\>_{e^{\dash6\phi}\!R^4}~\neq~\bigl\<\varphi^{12|56}\varphi^{34|78}++--\bigr\>_{e^{\dash6\phi}\!R^4}~\neq~
    -\bigl\<\varphi^{123|5}\varphi^{4|678}++--\bigr\>_{e^{\dash6\phi}\!R^4}  \,.
\end{equation}

The entire $SU(8)$ violation can be attributed to the local operator $\phi^2 R^4$. Its $SU(8)$-violating local matrix elements are encoded in a superamplitude proportional
  to 
${\cal O}_6^{\rm NMHV}$ in~(\ref{O6NMHVsuper}). Its only non-vanishing basis matrix elements are the 3 scalar ones with the same external states as those in \reef{scffR4}. The idea is now to relate the $6$-point NMHV superamplitudes of $e^{\dash6\phi}\!R^4$ and $R^4$ as
\bea
  \lab{opreq}
   {\cal M}^\text{NMHV}_{6,\,e^{\dash6\phi}\!R^4} ~=~
  {\cal M}^\text{NMHV}_{6,\,R^4}
  +  c\, {\cal O}_6^{\rm NMHV} \,.
\eea
The left-hand side will be obtained from the $\a'$-expansion of the string amplitude, with the normalization of $e^{-6\phi}R^4$ specified in~(\ref{norm}). Our goal is to determine the constant $c$ such that we can isolate the $SU(8)$-invariant ${\cal M}^\text{NMHV}_{6,\,R^4}$.

We use KLT to calculate the basis amplitudes of ${\cal M}^\text{NMHV}_{6,\,e^{\dash6\phi}\!R^4}$. In particular, we find 
\begin{equation}
  \label{diff}
    \bigl\<\varphi^{1234}\varphi^{5678}++--\bigr\>_{e^{\dash6\phi}\!R^4}~-\bigl\<\varphi^{12|56}\varphi^{34|78}++--\bigr\>_{e^{\dash6\phi}\!R^4}~=~6\times [34]^4\<56\>^4\,.
\end{equation}
The $SU(8)$-invariant contributions cancel in this difference, since the external states are $SU(8)$-equivalent. Thus the RHS of \reef{diff} is local and it comes entirely from the unique $SU(8)$-violating operator ${\cal O}_6^{\rm NMHV}$ \reef{O6NMHVsuper}. We found in (\ref{O6basis}) that locality of the matrix elements of ${\cal O}_6^{\rm NMHV}$ required the difference of the two relevant scalar basis elements to be $12-2=10$. Thus \reef{opreq} and \reef{diff} tells us that $10c=6$, i.e.~$c=3/5$. Then~\reef{opreq} gives
\bea
  \lab{opreq2}
  {\cal M}^\text{NMHV}_{6,\,R^4} ~=~
  {\cal M}^\text{NMHV}_{6,\,e^{\dash6\phi}\!R^4} -   \frac{3}{5}\, {\cal O}_6^{\rm NMHV} \,.
\eea
We have explicitly checked that the basis amplitudes resulting from \reef{opreq2} satisfy all $SU(8)$ conditions of~(\ref{SU8cond}), and hence all matrix elements computed from \reef{opreq2} satisfy the $SU(8)$ Ward identities. In particular we have
\begin{equation}
    \bigl\<\varphi^{1234}\varphi^{5678}++--\bigr\>_{\!R^4}~=~
    \bigl\<\varphi^{12|56}\varphi^{34|78}++--\bigr\>_{\!R^4}~=~-\bigl\<\varphi^{123|5}\varphi^{4|678}++--\bigr\>_{\!R^4}  \,,
\end{equation}
in contrast to the matrix elements to $e^{-6\phi}R^4$.

\subsection{Soft limits}

We can now use the relationship \reef{opreq2} to examine the soft scalar limits of 6-point  matrix elements of $R^4$. As in~(\ref{softp3}), the string amplitudes give us
\begin{equation}
    \lim_{p_1\to 0}\bigl\<\varphi^{1234} \varphi^{5678}++--\bigr\>_{e^{\dash6\phi}\!R^4}=6\times[34]^4\<56\>^4\,,  \quad
    \lim_{p_1\to 0}\bigl\<\varphi^{123|5} \varphi^{4|678}++--\bigr\>_{e^{\dash6\phi}\!R^4}=3\times[34]^4\<56\>^4\,,
\end{equation}
while the soft-limit of the 
 third 
scalar basis amplitude vanishes:
\begin{equation}
    \lim_{p_1\to 0}\bigl\<\varphi^{12|56} \varphi^{34|78}++--\bigr\>_{e^{\dash6\phi}\!R^4}~=~0\,.
\end{equation}

The corresponding soft limits of $R^4$ are identified by subtracting the $SU(8)$-violating contribution appropriately according to~(\ref{opreq2}). We find
\begin{equation}
    \lim_{p_1\to 0}\bigl\<\varphi^{1234}\, \varphi ^{5678}\,++--\bigr\>_{\!R^4}~\,=~-\frac{6}{5}\,[34]^4\<56\>^4\,,
\end{equation}
as well as identical soft limits for all the $SU(8)$-equivalent matrix elements.
This agrees with the result~(\ref{soft2}) found in section \ref{s:SSLgrp} by averaging the $\a'^3$-contributions of the closed string amplitude over $SU(8)$.

In~\cite{BD} it was pointed out that the $6$-point NMHV matrix elements of $e^{-6\phi}R^4$, despite their non-vanishing {\em single}-soft scalar limits, behave under {\em double}-soft scalar limits as would be required for an $E_{7(7)}$-invariant operator~\cite{ArkaniHamed:2008gz}. The double-soft scalar limits reveal the group structure of the broken $E_{7(7)}$ generators through the non-commutativity of soft limits of two distinct scalar lines. The double-soft limit analysis is only sensitive to non-local terms, because soft limits on local terms commute. The NMHV $6$-point matrix elements of $R^4$ and $e^{-6\phi}R^4$ differ only by local terms. Therefore, the double-soft scalar limits of 
 the $6$-point matrix elements of 
$R^4$, curiously, also satisfy the consistency relations of~\cite{ArkaniHamed:2008gz}, even though the single-soft scalar limits have revealed that it does not respect $E_{7(7)}$.

\setcounter{equation}{0}
\section{Linear relations between $F^4$ matrix elements}\label{s:vanishing}
The intriguing reformulation of the KLT relations as the BCJ squaring relations~\cite{BCJ} has recently lead to a renewed interest in the study of `gravity=(gauge theory)$^2$'. The BCJ squaring relations have been extended to loop level~\cite{BCJloop,BDHK}, and have motivated the construction of variations of  KLT~\cite{KLTvar}. In~\cite{BDFS}, BCFW recursion relations~\cite{bcfw} were used to derive KLT-like  `0=(gauge theory)$^2$' relations. As also pointed out in~\cite{Tye}, these ``vanishing relations'' can be understood from the observation that
N$^{(k,\tk)}$MHV amplitudes in ${\cal N}=8$ supergravity vanish for $k\neq \tilde{k}$. The KLT relations then give quadratic relations 
between ${\cn=4}$ SYM amplitudes of distinct N$^k$MHV sectors~\cite{Tye,Feng}.

A particularly 
 simple 
from of these identities can be obtained from setting $\tk=0$, and explicitly plugging in the Parke-Taylor formula for  the MHV amplitudes $\tilde{A}^{\rm MHV}_n$. As was also pointed out in~\cite{Feng}, one then obtains linear relations between N$^k$MHV amplitudes, for any $k>0$.
For example, at 5-point, KLT then gives the following relation between $5$-point NMHV (=$\rm \overline{MHV}$) amplitudes of ${\cn=4}$ SYM:
\begin{equation}
\begin{split}
  0~=~
   -  [12][34] A_5^{\rm \overline{MHV}}(2,1,4,3,5) + [13][24] A_5^{\rm \overline{MHV}}(3,1,4,2,5)\,.
\end{split}
\end{equation}
This relation is of course immediately obvious from the conjugated Parke-Taylor formula. For $n>5$, however, the analogous relations are 
 non-trivial. 
For example, at $6$-point, they give linear relations that are valid both for NMHV and anti-MHV $6$-point amplitudes. Instead of exploring these identities for higher-point amplitudes, we now point out similar identities for the matrix elements of the $F^4$ operator in the open-string effective action.

The leading $\alpha'$-correction  in the closed string effective action occurs at order $\alpha'^3$. As was also pointed out in~\cite{BD}, closed string tree amplitudes therefore receive no contributions at order $\alpha'^2$:
\begin{equation}
    M_n^{(k,\tk)}\Bigr|_{O(\alpha'^2)}~=~0\qquad \qquad\text{(for any $k,\tk$)}\,.
\end{equation}
 The leading correction to the open string tree amplitude, however, is already at order $\alpha'^2$. In the effective action, it is captured by the operator $\zeta(2)F^4$. Therefore, KLT implies {\em linear} relations between the $n$-point matrix elements of $F^4$. The coefficients of these linear relations are functions of Mandelstam variables, multiplied by tree-amplitudes of ${\cal N}=4$ SYM. Schematically, we obtain
\begin{equation}\label{genvanish}
    \sum_\mathcal{P} \Bigl[
    f^{(2)}_n\,A_{\rm SYM} \times \tilde{A}_{\rm SYM}
    \,+\,f^{(0)}_n\,A_{\rm SYM} \times \tilde{A}_{F^4}
    \,+\,f^{(0)}_n\,A_{F^4} \times \tilde{A}_{\rm SYM}
    \Bigr]\,=\,0\,,
\end{equation}
where $f^{(i)}_n$ are functions of the Mandelstam variables $s_{ij}$, constructed from the $O(\alpha'^i)$ contribution to $f_n$ in~(\ref{KLTschematic}). Note that $A$ and $\tilde{A}$ in~(\ref{genvanish}) can belong to any sector N$^k$MHV and N$^{\tk}$MHV, respectively. We do {\em not} have to demand $k\neq \tk$ as for the leading $\alpha'\to0$ identities.

We now illustrate these identities for the $5$-point case. We obtain
\begin{equation}
\begin{split}
  0~=~s_{12}s_{34}\Bigl[&A_{ F^4}(1,2,3,4,5)\widetilde{A}_{\rm SYM}(2,1,4,3,5)+A_{\rm SYM}(1,2,3,4,5)\widetilde{A}_{F^4}(2,1,4,3,5)\\[1ex]
  &-\bigl(s_{12}^2+s_{34}^2)A_{\rm SYM}(1,2,3,4,5)\widetilde{A}_{\rm SYM}(2,1,4,3,5)\Bigr]
  ~+~ (2\lra 3)\,,
\end{split}
\end{equation}
where we normalized the operator $F^4$ such that it appears as $\zeta(2)\alpha'^2F^4$ in the open string effective action. Similar identities can be derived for $n>5$ to relate higher-point non-local matrix elements of $F^4$ with different color-orderings.

Before closing this section, let us point out two obvious generalizations of the stringy vanishing identities we introduced above:
\begin{enumerate}
  \item[i.] The closed string tree amplitude with massless external states also vanishes at order $\alpha'^4$. This leads to {\em quadratic} vanishing identities involving both tree amplitudes of $\cn=4$ SYM and matrix elements of $F^4$.
  \item[ii.] As the closed string tree amplitude vanishes at order $\alpha'^2$ and $\alpha'^4$ in any dimension~\cite{StStGravity}, the vanishing relations for $F^4$ are valid in any number of dimensions. 
      This is to be contrasted with the vanishing relations for tree amplitudes, which are intrinsically 4-dimensional because they rely on an N$^k$MHV classification for open string amplitudes. 
       Perhaps it is possible to find an analogous classification of higher-dimensional amplitudes to also extend the tree-level vanishing relations to $D>4$.
\end{enumerate}
While these identities between gauge-theory matrix elements emerge naturally from KLT, it would be interesting to elucidate their origin and role within $\cn=4$ SYM theory.

 \section*{Acknowledgements}
We are indebted to Juan Maldacena and Edward Witten for very helpful discussions and suggestions. We are grateful to Stephan Stieberger and Tomasz Taylor for generously providing us with the Mathematica code for the open string tree amplitudes.
 We thank Niklas Beisert, Daniel Freedman, and Alejandro Morales for collaboration on related topics, and also Emil Bjerrum-Bohr, Johannes Br\"odel, Poul Damgaard, Lance Dixon, Michael Green, Nathan Seiberg and Yuji Tachikawa for valuable discussions.

HE is supported in part by the US Department of Energy under DOE grants DE-FG02-95ER40899 (Michigan) and DE-FG02-90ER40542 (IAS).
The research of MK is supported by NSF grant PHY-0756966.


\begin{thebibliography}{99}

%\cite{Kawai:1985xq}
\bibitem{KLT}
  H.~Kawai, D.~C.~Lewellen and S.~H.~H.~Tye,
  ``A Relation Between Tree Amplitudes Of Closed And Open Strings,''
  Nucl.\ Phys.\  B {\bf 269}, 1 (1986).
  %%CITATION = NUPHA,B269,1;%%




\bibitem{BernFiniteness}
%\cite{Bern:2007hh}
%\bibitem{Bern:2007hh}
  Z.~Bern, J.~J.~Carrasco, L.~J.~Dixon, H.~Johansson, D.~A.~Kosower and R.~Roiban,
  ``Three-Loop Superfiniteness of N=8 Supergravity,''
  Phys.\ Rev.\ Lett.\  {\bf 98}, 161303 (2007)
  [arXiv:hep-th/0702112].
  %%CITATION = PRLTA,98,161303;%%

%\cite{Bern:2008pv}
%\bibitem{Bern:2008pv}
  Z.~Bern, J.~J.~M.~Carrasco, L.~J.~Dixon, H.~Johansson and R.~Roiban,
  ``Manifest Ultraviolet Behavior for the Three-Loop Four-Point Amplitude of
  N=8 Supergravity,''
  Phys.\ Rev.\  D {\bf 78}, 105019 (2008)
  [arXiv:0808.4112 [hep-th]].
  %%CITATION = PHRVA,D78,105019;%%

  %\cite{Bern:2009kf}
%\bibitem{Bern:2009kf}
  Z.~Bern, J.~J.~M.~Carrasco and H.~Johansson,
  ``Progress on Ultraviolet Finiteness of Supergravity,''
  arXiv:0902.3765 [hep-th].
  %%CITATION = ARXIV:0902.3765;%%

%\cite{Bern:2009kd}
%\bibitem{Bern:2009kd}
  Z.~Bern, J.~J.~Carrasco, L.~J.~Dixon, H.~Johansson and R.~Roiban,
  ``The Ultraviolet Behavior of N=8 Supergravity at Four Loops,''
  Phys.\ Rev.\ Lett.\  {\bf 103}, 081301 (2009)
  [arXiv:0905.2326 [hep-th]].
  %%CITATION = PRLTA,103,081301;%%


\bibitem{Unitarity}
Z.~Bern, L.~J.~Dixon, D.~C.~Dunbar and D.~A.~Kosower,
``One loop n point gauge theory amplitudes, unitarity and collinear limits,''
Nucl.\ Phys.\ B {\bf 425}, 217 (1994)
[arXiv:hep-ph/9403226];
%%CITATION = HEP-PH 9403226;%%

Z.~Bern, L.~J.~Dixon, D.~C.~Dunbar and D.~A.~Kosower,
``Fusing gauge theory tree amplitudes into loop amplitudes,''
Nucl.\ Phys.\ B {\bf 435}, 59 (1995)
[arXiv:hep-ph/9409265];
%%CITATION = HEP-PH 9409265;%%

 Z.~Bern, L.~J.~Dixon and D.~A.~Kosower,
  ``Two-Loop g $\to$ gg Splitting Amplitudes in QCD,''
  JHEP {\bf 0408}, 012 (2004)
  [arXiv:hep-ph/0404293].
  %%CITATION = JHEPA,0408,012;%%


\bibitem{Counterterms}

%\cite{Howe:1980th}
%\bibitem{Howe:1980th}
  P.~S.~Howe and U.~Lindstrom,
  ``Higher Order Invariants In Extended Supergravity,''
  Nucl.\ Phys.\  B {\bf 181}, 487 (1981).
  %%CITATION = NUPHA,B181,487;%%

%\cite{Kallosh:1980fi}
%\bibitem{Kallosh:1980fi}
  R.~E.~Kallosh,
  ``Counterterms in extended supergravities,''
  Phys.\ Lett.\  B {\bf 99}, 122 (1981).
  %%CITATION = PHLTA,B99,122;%%



%\cite{Howe:2002ui}
%\bibitem{Howe:2002ui}
  P.~S.~Howe and K.~S.~Stelle,
  ``Supersymmetry counterterms revisited,''
  Phys.\ Lett.\  B {\bf 554}, 190 (2003)
  [arXiv:hep-th/0211279].
  %%CITATION = PHLTA,B554,190;%%


%\cite{Stelle:2007zz}
%\bibitem{Stelle:2007zz}
  K.~Stelle,
  ``Finite after all?,''
  Nature Phys.\  {\bf 3}, 448 (2007).
  %%CITATION = NPAHA,3,448;%%

%\cite{Bossard:2009sy}
%\bibitem{Bossard:2009sy}
  G.~Bossard, P.~S.~Howe and K.~S.~Stelle,
  ``The ultra-violet question in maximally supersymmetric field theories,''
  Gen.\ Rel.\ Grav.\  {\bf 41}, 919 (2009)
  [arXiv:0901.4661 [hep-th]].
  %%CITATION = GRGVA,41,919;%%

%\cite{Bossard:2009mn}
%\bibitem{Bossard:2009mn}
  G.~Bossard, P.~S.~Howe and K.~S.~Stelle,
  ``A note on the UV behaviour of maximally supersymmetric Yang-Mills
  theories,''
  Phys.\ Lett.\  B {\bf 682}, 137 (2009)
  [arXiv:0908.3883 [hep-th]].
  %%CITATION = PHLTA,B682,137;%%

 %\cite{Stelle:2009zz}
%\bibitem{Stelle:2009zz}
  K.~S.~Stelle,
  ``Is N=8 supergravity a finite field theory?,''
  Fortsch.\ Phys.\  {\bf 57}, 446 (2009).
  %%CITATION = FPYKA,57,446;%%

\bibitem{Harmonic}

 %\cite{Drummond:2003ex}
%\bibitem{Drummond:2003ex}
  J.~M.~Drummond, P.~J.~Heslop, P.~S.~Howe and S.~F.~Kerstan,
  ``Integral invariants in N = 4 SYM and the effective action for  coincident
  D-branes,''
  JHEP {\bf 0308}, 016 (2003)
  [arXiv:hep-th/0305202].
  %%CITATION = JHEPA,0308,016;%%


\bibitem{StringyCounterterms}

%\cite{Berkovits:2006vc}
%\bibitem{Berkovits:2006vc}
  N.~Berkovits,
  ``New higher-derivative R**4 theorems,''
  Phys.\ Rev.\ Lett.\  {\bf 98}, 211601 (2007)
  [arXiv:hep-th/0609006].
  %%CITATION = PRLTA,98,211601;%%

%\cite{Green:2006gt}
%\bibitem{Green:2006gt}
  M.~B.~Green, J.~G.~Russo and P.~Vanhove,
  ``Non-renormalisation conditions in type II string theory and maximal
  supergravity,''
  JHEP {\bf 0702}, 099 (2007)
  [arXiv:hep-th/0610299].
  %%CITATION = JHEPA,0702,099;%%

  %\cite{Green:2006yu}
%\bibitem{Green:2006yu}
  M.~B.~Green, J.~G.~Russo and P.~Vanhove,
  ``Ultraviolet properties of maximal supergravity,''
  Phys.\ Rev.\ Lett.\  {\bf 98}, 131602 (2007)
  [arXiv:hep-th/0611273].
  %%CITATION = PRLTA,98,131602;%%


%\cite{Green:2008bf}
%\bibitem{Green:2008bf}
  M.~B.~Green, J.~G.~Russo and P.~Vanhove,
  ``Modular properties of two-loop maximal supergravity and connections with
  string theory,''
  JHEP {\bf 0807}, 126 (2008)
  [arXiv:0807.0389 [hep-th]].
  %%CITATION = JHEPA,0807,126;%%


%\cite{Berkovits:2009aw}
%\bibitem{Berkovits:2009aw}
  N.~Berkovits, M.~B.~Green, J.~G.~Russo and P.~Vanhove,
  ``Non-renormalization conditions for four-gluon scattering in supersymmetric
  string and field theory,''
  JHEP {\bf 0911}, 063 (2009)
  [arXiv:0908.1923 [hep-th]].
  %%CITATION = JHEPA,0911,063;%%



%\cite{Vanhove:2010nf}
%\bibitem{Vanhove:2010nf}
  P.~Vanhove,
  ``The critical ultraviolet behaviour of N=8 supergravity amplitudes,''
  arXiv:1004.1392 [hep-th].
  %%CITATION = ARXIV:1004.1392;%%



\bibitem{Kallosh}

%\cite{Kallosh:2008mq}
%\bibitem{Kallosh:2008mq}
  R.~Kallosh,
  ``On a possibility of a UV finite N=8 supergravity,''
  arXiv:0808.2310 [hep-th].
  %%CITATION = ARXIV:0808.2310;%%


%\cite{Kallosh:2009db}
%\bibitem{Kallosh:2009db}
  R.~Kallosh,
  ``N=8 Supergravity on the Light Cone,''
  Phys.\ Rev.\  D {\bf 80}, 105022 (2009)
  [arXiv:0903.4630 [hep-th]].
  %%CITATION = PHRVA,D80,105022;%%


%\cite{Kallosh:2009jb}
%\bibitem{Kallosh:2009jb}
  R.~Kallosh,
  ``On UV Finiteness of the Four Loop N=8 Supergravity,''
  JHEP {\bf 0909}, 116 (2009)
  [arXiv:0906.3495 [hep-th]].
  %%CITATION = JHEPA,0909,116;%%

\bibitem{Eisenstein}

%\cite{Green:2010wi}
%\bibitem{Green:2010wi}
  M.~B.~Green, J.~G.~Russo and P.~Vanhove,
  ``Automorphic properties of low energy string amplitudes in various
  dimensions,''
  arXiv:1001.2535 [hep-th].
  %%CITATION = ARXIV:1001.2535;%%
  
%\cite{Green:2010sp}
%\bibitem{Green:2010sp}
  M.~B.~Green, J.~G.~Russo and P.~Vanhove,
  ``String theory dualities and supergravity divergences,''
  JHEP {\bf 1006}, 075 (2010)
  [arXiv:1002.3805 [hep-th]].
  %%CITATION = JHEPA,1006,075;%%


%\cite{Green:2010kv}
%\bibitem{Green:2010kv}
  M.~B.~Green, S.~D.~Miller, J.~G.~Russo and P.~Vanhove,
  ``Eisenstein series for higher-rank groups and string theory amplitudes,''
  arXiv:1004.0163 [hep-th].
  %%CITATION = ARXIV:1004.0163;%%




%\cite{Elvang:2010jv}
\bibitem{efk5}
  H.~Elvang, D.~Z.~Freedman and M.~Kiermaier,
  ``A simple approach to counterterms in N=8 supergravity,''
  arXiv:1003.5018 [hep-th].
  %%CITATION = ARXIV:1003.5018;%%


%\cite{Dixon:2010gz}
\bibitem{LanceReview}
  L.~J.~Dixon,
  ``Ultraviolet Behavior of N=8 Supergravity,''
  arXiv:1005.2703 [hep-th].
  %%CITATION = ARXIV:1005.2703;%%


\bibitem{N8}


%\cite{Cremmer:1978ds}
%\bibitem{Cremmer:1978ds}
  E.~Cremmer and B.~Julia,
  ``The N=8 Supergravity Theory. 1. The Lagrangian,''
  Phys.\ Lett.\  B {\bf 80}, 48 (1978).
  %%CITATION = PHLTA,B80,48;%%

%\cite{Cremmer:1979up}
%\bibitem{Cremmer:1979up}
  E.~Cremmer and B.~Julia,
  ``The SO(8) Supergravity,''
  Nucl.\ Phys.\  B {\bf 159}, 141 (1979).
  %%CITATION = NUPHA,B159,141;%%

%\cite{Bianchi:2008pu}
\bibitem{BEF}
  M.~Bianchi, H.~Elvang and D.~Z.~Freedman,
  ``Generating Tree Amplitudes in N=4 SYM and N = 8 SG,''
  JHEP {\bf 0809}, 063 (2008)
  [arXiv:0805.0757 [hep-th]].
  %%CITATION = JHEPA,0809,063;%%


%\cite{ArkaniHamed:2008gz}
\bibitem{ArkaniHamed:2008gz}
  N.~Arkani-Hamed, F.~Cachazo and J.~Kaplan,
  ``What is the Simplest Quantum Field Theory?,''
  arXiv:0808.1446 [hep-th].
  %%CITATION = ARXIV:0808.1446;%%


\bibitem{footprint}
%\cite{Kallosh:2008rr}
%\bibitem{Kallosh:2008rr}
  R.~Kallosh and T.~Kugo,
  ``The footprint of E7 in amplitudes of N=8 supergravity,''
  JHEP {\bf 0901}, 072 (2009)
  [arXiv:0811.3414 [hep-th]].
  %%CITATION = JHEPA,0901,072;%%



\bibitem{STMHV}

%\cite{Stieberger:2006te}
%\bibitem{Stieberger:2006te}
  S.~Stieberger and T.~R.~Taylor,
  ``Multi-gluon scattering in open superstring theory,''
  Phys.\ Rev.\  D {\bf 74}, 126007 (2006)
  [arXiv:hep-th/0609175].
  %%CITATION = PHRVA,D74,126007;%%


\bibitem{ST6pt}

%\cite{Stieberger:2007am}
%\bibitem{Stieberger:2007am}
  S.~Stieberger and T.~R.~Taylor,
  ``Complete Six-Gluon Disk Amplitude in Superstring Theory,''
  Nucl.\ Phys.\  B {\bf 801}, 128 (2008)
  [arXiv:0711.4354 [hep-th]].
  %%CITATION = NUPHA,B801,128;%%


%\cite{Broedel:2009hu}
\bibitem{BD}
  J.~Broedel and L.~J.~Dixon,
  ``$R^4$ counterterm and E7(7) symmetry in maximal supergravity,''
  arXiv:0911.5704 [hep-th].
  %%CITATION = ARXIV:0911.5704;%%


%\cite{Elvang:2009wd}
\bibitem{efk4}
  H.~Elvang, D.~Z.~Freedman and M.~Kiermaier,
  ``Solution to the Ward Identities for Superamplitudes,''
  arXiv:0911.3169 [hep-th].
  %%CITATION = ARXIV:0911.3169;%%




\bibitem{bcfw}
%\cite{Britto:2004ap}
%\bibitem{Britto:2004ap}
  R.~Britto, F.~Cachazo and B.~Feng,
  ``New Recursion Relations for Tree Amplitudes of Gluons,''
  Nucl.\ Phys.\  B {\bf 715}, 499 (2005)
  [arXiv:hep-th/0412308].
  %%CITATION = NUPHA,B715,499;%%

 R.~Britto, F.~Cachazo, B.~Feng and E.~Witten,
  ``Direct Proof Of Tree-Level Recursion Relation In Yang-Mills Theory,''
  Phys.\ Rev.\ Lett.\  {\bf 94}, 181602 (2005)
  [arXiv:hep-th/0501052].
  %%CITATION = PRLTA,94,181602;%%

  \bibitem{BDFS}

%\cite{BjerrumBohr:2010zb}
%\bibitem{BjerrumBohr:2010zb}
  N.~E.~J.~Bjerrum-Bohr, P.~H.~Damgaard, B.~Feng and T.~Sondergaard,
  ``New Identities among Gauge Theory Amplitudes,''
  arXiv:1006.3214 [hep-th].
  %%CITATION = ARXIV:1006.3214;%%

%\cite{BjerrumBohr:2010yc}
%\bibitem{BjerrumBohr:2010yc}
  N.~E.~J.~Bjerrum-Bohr, P.~H.~Damgaard, B.~Feng and T.~Sondergaard,
  ``Proof of Gravity and Yang-Mills Amplitude Relations,''
  arXiv:1007.3111 [hep-th].
  %%CITATION = ARXIV:1007.3111;%%



%\cite{Feng:2010br}
\bibitem{Feng}
  B.~Feng and S.~He,
  ``KLT and New Relations for N=8 SUGRA and N=4 SYM,''
  arXiv:1007.0055 [hep-th].
  %%CITATION = ARXIV:1007.0055;%%

%\cite{Tye:2010kg}
\bibitem{Tye}
  H.~Tye and Y.~Zhang,
  ``Comment on the Identities of the Gluon Tree Amplitudes,''
  arXiv:1007.0597 [hep-th].
  %%CITATION = ARXIV:1007.0597;%%


\bibitem{STWI}

%\cite{Stieberger:2007jv}
%\bibitem{Stieberger:2007jv}
  S.~Stieberger and T.~R.~Taylor,
  ``Supersymmetry Relations and MHV Amplitudes in Superstring Theory,''
  Nucl.\ Phys.\  B {\bf 793}, 83 (2008)
  [arXiv:0708.0574 [hep-th]].
  %%CITATION = NUPHA,B793,83;%%



\bibitem{csw}
  F.~Cachazo, P.~Svrcek and E.~Witten,
  ``MHV vertices and tree amplitudes in gauge theory,''
  JHEP {\bf 0409}, 006 (2004)
  [arXiv:hep-th/0403047].
  %%CITATION = JHEPA,0409%\cite{Britto:2005fq}







\bibitem{Pioline}
%\cite{Obers:1999es}
%\bibitem{Obers:1999es}
  N.~A.~Obers and B.~Pioline,
  ``Eisenstein series in string theory,''
  Class.\ Quant.\ Grav.\  {\bf 17}, 1215 (2000)
  [arXiv:hep-th/9910115].
  %%CITATION = CQGRD,17,1215;%%

%\cite{Pioline:2010kb}
%\bibitem{Pioline:2010kb}
  B.~Pioline,
  ``R**4 couplings and automorphic unipotent representations,''
  JHEP {\bf 1003}, 116 (2010)
  [arXiv:1001.3647 [hep-th]].
  %%CITATION = JHEPA,1003,116;%%

%he ref added
\bibitem{future}
%\cite{Beisert:2010jx}
%\bibitem{Beisert:2010jx}
  N.~Beisert, H.~Elvang, D.~Z.~Freedman, M.~Kiermaier, A.~Morales and S.~Stieberger,
  ``E7(7) constraints on counterterms in N=8 supergravity,''
  arXiv:1009.1643 [hep-th].
  %%CITATION = ARXIV:1009.1643;%%


\bibitem{BCJ}
%\cite{Bern:2008qj}
%\bibitem{Bern:2008qj}
  Z.~Bern, J.~J.~M.~Carrasco and H.~Johansson,
  ``New Relations for Gauge-Theory Amplitudes,''
  Phys.\ Rev.\  D {\bf 78}, 085011 (2008)
  [arXiv:0805.3993 [hep-ph]].
  %%CITATION = PHRVA,D78,085011;%%

\bibitem{BCJloop}
%\cite{Bern:2010ue}
%\bibitem{Bern:2010ue}
  Z.~Bern, J.~J.~M.~Carrasco and H.~Johansson,
  ``Perturbative Quantum Gravity from Gauge Theory,''
  arXiv:1004.0476 [hep-th].
  %%CITATION = ARXIV:1004.0476;%%

\bibitem{BDHK}
%\cite{Bern:2010yg}
%\bibitem{Bern:2010yg}
  Z.~Bern, T.~Dennen, Y.~t.~Huang and M.~Kiermaier,
  ``Gravity as the Square of Gauge Theory,''
  arXiv:1004.0693 [hep-th].
  %%CITATION = ARXIV:1004.0693;%%


\bibitem{KLTvar}

%\cite{BjerrumBohr:2010ta}
%\bibitem{BjerrumBohr:2010ta}
  N.~E.~J.~Bjerrum-Bohr, P.~H.~Damgaard, B.~Feng and T.~Sondergaard,
  ``Gravity and Yang-Mills Amplitude Relations,''
  arXiv:1005.4367 [hep-th].
  %%CITATION = ARXIV:1005.4367;%%


\bibitem{StStGravity}
%\cite{Stieberger:2009rr}
%\bibitem{Stieberger:2009rr}
  S.~Stieberger,
  ``On tree-level higher order gravitational couplings in superstring theory,''
  arXiv:0910.0180 [hep-th].
  %%CITATION = ARXIV:0910.0180;%%







\end{thebibliography}
\end{document}